\documentclass[prd,aps,preprint,nofootinbib,superscriptaddress,tightenlines]{revtex4}
\pdfoutput=1
\usepackage[english]{babel}
\usepackage[latin1]{inputenc}
\usepackage{graphicx}
\usepackage{amsmath}
\usepackage{amssymb}
\usepackage{xcolor}
\usepackage{hyperref}
\usepackage{subfigure}
\usepackage{units}

\newcommand{\etal}{\textit{et~al.}}
\newcommand{\bs}[1]{\boldsymbol{#1}}
\newcommand{\XX}{$X(3872)$}
\newcommand{\X}{$X$}
\newcommand{\Ecm}{E_\text{cm}}
\newcommand{\KLO}{K_\text{lo}}
\newcommand{\KHI}{K_\text{hi}}
\newcommand{\dX}{\delta_X}
\newcommand{\GDPi}{\Gamma_{\!D\pi}}
\newcommand{\GDg}{\Gamma_{\!D\gamma}}
\newcommand{\Gast}{\Gamma_{\!\ast}}
\newcommand{\Gc}{\Gamma_{\!\text{c}}}
\newcommand{\GX}{\Gamma_{\!X}}

\begin{document}

\title{Threshold Effects and the Line Shape of the $\bs{X(3872)}$
  in Effective Field Theory}
\author{M. Schmidt} \affiliation{Institut f\"ur Kernphysik, Technische Universit\"at Darmstadt, 64289 Darmstadt, Germany
}
\author{M. Jansen} \affiliation{Institut f\"ur Kernphysik, Technische Universit\"at Darmstadt, 64289 Darmstadt, Germany
}
\author{H.-W. Hammer} \affiliation{Institut f\"ur Kernphysik, Technische Universit\"at Darmstadt, 64289 Darmstadt, Germany
}
\affiliation{ExtreMe Matter Institute EMMI, GSI Helmholtzzentrum f\"ur Schwerionenforschung GmbH,
64291 Darmstadt, Germany
}
\date{\today}

\begin{abstract}
Latest measurements suggest that the \XX\ mass lies less than $\unit[200]{keV}$
away from the $D^0\bar{D}^{0\ast}$ threshold, reenforcing its interpretation as
a loosely-bound mesonic molecule. This observation implies that in processes
like $D^0\bar{D}^0\pi^0$ production, threshold effects could disguise the actual
pole position of the \XX.
We propose a new effective field theory with $D^0$, $\bar{D}^0$ and $\pi^0$ degrees
of freedom for the \XX, considering Galilean invariance to be an exact symmetry.
The $D^{0\ast}$ enters as a $D^0\pi^0$ $p$-wave resonance, allowing for a
comprehensive study of the influence of pion interactions on the \XX\ width.
We calculate relations between the mass of the \XX, its width, and its line shape
in $D^0\bar{D}^0\pi^0$ production up to next-to-leading order. Our results
provide a tool for the extraction of the \XX\ pole position from the experimental
data near threshold.
\end{abstract}

\smallskip
\maketitle

\section{Introduction}

In 2003 and 2004, the Belle and the CDF II collaborations subsequently observed a novel charmonium state referred to as the \XX\ \cite{Choi:2003ue, Acosta:2003zx}. Since then, experimentalists have discovered a whole zoo of exotic ``$XYZ$ particles'' that do not fit into the conventional quark model. So far, their nature remains unclear and is under discussion (see, e.g., Refs.~\cite{Godfrey:2008nc,Hosaka:2016pey,Shen:2017kom,Lebed:2016hpi} for reviews). The \XX\ (simply denoted '\X' in the following) is by far the best-studied example of these states. Its decay channels $J/\psi\,\pi^+\pi^-$ and $J/\psi\,\pi^+\pi^-\pi^0$ have comparable branching ratios \cite{Choi:2011fc, delAmoSanchez:2010vq}, implying a large isospin violation. For this reason, its interpretation as a conventional $c\bar{c}$ state has been challenged by a variety of exotic explanations including an interpretation as a tetraquark~\cite{Terasaki:2007uv, Terasaki:2016yvc}. A major step towards a deeper understanding of the \X\ was achieved in 2013 when the LHCb Collaboration determined its quantum numbers to be $J^{PC}=1^{++}$ \cite{Aaij:2013zoa}. For a recent review on the experimental status of the \X, see Ref.~\cite{Aushev:2016erz}.

\begin{figure}
	\includegraphics[scale=1]{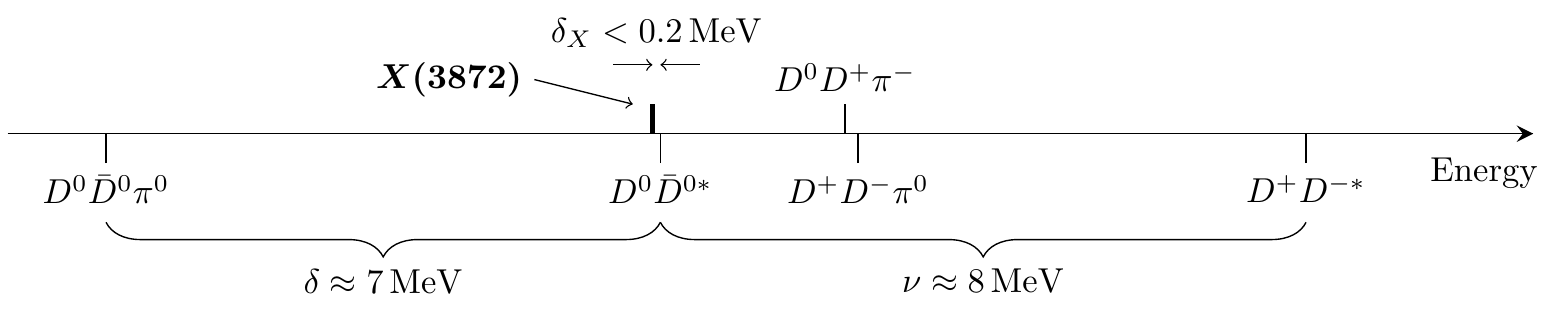}
	\vspace{-3mm}
	\caption{\label{Fig:MassSketch}Thresholds close to the \XX\ (to scale, anti-particle states omitted).}
\end{figure}

A striking feature of the \X\ state is its proximity to
the $D^0\bar{D}^{0\ast}$ threshold; see Fig.~\ref{Fig:MassSketch}. Over the years, the corresponding mass difference has repeatedly been corrected down from about $\unit[+1]{MeV}$ to the current value \cite{Tomaradze:2015cza, PDG2017}
\begin{equation}
	\label{Eq:dX}
	\dX\equiv \left(
		m_{D^0}+m_{D^{0\ast}}
	\right)
	-m_X =\unit[(-0.01\pm 0.2)]{MeV}\,,
\end{equation}
with $m_X=\unit[3871.69(17)]{MeV}$. Considering the quantum numbers of the $X$,
$J^{PC}=1^{++}$, this circumstance implies a strong coupling to the $s$-wave configuration
$\left(D^0 \bar{D}^{0\ast}+\bar{D}^0 D^{0\ast}\right)/\sqrt{2}$, giving rise to a large molecular component. This picture, which has been discussed by many authors
(cf.~Refs.~\cite{Close:2003sg, Pakvasa:2003ea, Voloshin:2003nt, Wong:2003xk, Braaten:2003he, Swanson:2003tb}), is the basis of this work. It readily explains the isospin violation from the remoteness of the charged $D^+ D^{-\ast}$ threshold at
\begin{equation}
	\label{Eq:DeltaM}
	\nu \equiv \left(
		m_{D^+}+m_{D^{+\ast}}
	\right)-\left(
		m_{D^0}+m_{D^{0\ast}}
	\right)
	= \unit[8.16(9)]{MeV}\,;
\end{equation}
see Fig.~\ref{Fig:MassSketch} and Ref.~\cite{PDG2017}. Moreover, a significant part of the \X\ width $\GX$ can be attributed to decays of the constituents $D^{0\ast}$ and $\bar{D}^{0\ast}$, which is confirmed by the large branching ratios of \X\ decays involving $D^0\bar{D}^{0\ast}$ \cite{PDG2017}.
By now, the width is only limited by an upper bound $\GX<\unit[1.2]{MeV}$ \cite{Choi:2011fc} stemming from the detector resolution. However, future experiments like Belle II~\cite{HerediadelaCruz:2016yqi}
and $\bar{\text{P}}$ANDA at FAIR \cite{Prencipe:2015cgg}
will be able to measure the width with much higher accuracy.

If interpreted as a charm meson pair, the \X\ can either be bound or virtual, which follows from the universal properties of near-threshold $s$-wave states \cite{Braaten:2009zz, Braaten:2004rn}. Heavy quark symmetry then implies the existence of a $B^0\bar{B}^{0\ast}$ molecule \cite{AlFiky:2005jd}.
In a zero-range approach, Braaten and Lu calculated line shapes of the \X\ for the bound and virtual cases in 2007 \cite{Braaten:2007dw}. The partial widths of inelastic decay channels like the discovery mode $J/\psi\,\pi^+\pi^-$ were neglected, assuming that they are small. They observed a significant enhancement in the $D^0\bar{D}^0\pi^0$ production rate close to the $D^0\bar{D}^{0\ast}$ threshold. This effect influences the peak position and width, such that they do not correspond to the actual pole position. This could be one reason why the \X\ mass in the $D^0\bar{D}^0\pi^0$ channel in Refs.~\cite{Gokhroo:2006bt, Adachi:2008sua, Aubert:2007rva} appears to be larger than in such involving $J/\psi$ \cite{PDG2017}. In fact, Braaten and
Stapleton~\cite{Stapleton:2009ey} pointed out that the main reason for this result lies in a false identification of the peak in the invariant $D^0\bar{D}^{0\ast}$ mass distribution in Refs.~\cite{Adachi:2008sua, Aubert:2007rva}
as the \X\ mass and width. 
 An analysis of the effect of different pole structures on
  the line shapes of the X(3872) in $B$ decays to $KJ/\psi\,\pi^+\pi^-$ and
  $KJ/\psi\,D^0\bar{D}^{0\ast}$ was carried out in Ref.~\cite{Kang:2016jxw}.

In 2007, Fleming \etal\ developed a non-relativistic effective field theory (EFT) for the \X, called XEFT \cite{Fleming:2007rp}. It can be used to calculate systematic corrections to universality. In addition to charm meson fields, XEFT also contains a field for the neutral pion $\pi^0$. Fleming \etal\ calculated the partial decay width $\Gamma\left[X(3872)\rightarrow D^0\bar{D}^0\pi^0\right]$ at next-to-leading order (NLO) in XEFT power counting. A key result was that pion exchanges can be treated in perturbation theory. XEFT has been applied to several processes \cite{Fleming:2008yn, Fleming:2011xa, Mehen:2011ds, Margaryan:2013tta, Jansen:2013cba}. It is, however, limited to NLO precision because renormalization requires an expansion 
in the mass ratio $(m_{\pi^0}/m_{D^0})^{1/2}\approx 0.27$. Braaten cured this problem by proposing a Galilean-invariant version of XEFT \cite{Braaten:2015tga}. This symmetry is motivated by the small mass difference
\begin{equation}
	\label{Eq:Delta}
	\delta\equiv m_{D^{0\ast}} - m_{D^0} - m_{\pi^0}=\unit[7.04(3)]{MeV}
\end{equation}
in the decay $D^{0\ast}\rightarrow D^0\pi^0$,
which implies approximate mass conservation; see Fig.~\ref{Fig:MassSketch} and Ref.~\cite{PDG2017}.

The influence of pion dynamics on $\GX$ was also investigated by Baru \etal\ in 2011 \cite{Baru:2011rs}. They performed a coupled channel calculation with both neutral and charged mesons, treating pions non-perturbatively. The \X\ was produced as a peak in the $D^0\bar{D}^0\pi^0$ production rate at $\dX\geq \unit[0.1]{MeV}$. In this region, threshold effects play a minor role and $\GX$ can be extracted in a Breit-Wigner fit. Their result agrees well with XEFT, thereby
confirming the perturbativeness of pions. Moreover, it was shown that a static pion approximation largely overestimates $\GX$, while charged mesons have a moderate effect.

This work builds upon the findings of Refs.~\cite{Braaten:2007dw, Fleming:2007rp, Braaten:2015tga} and \cite{Baru:2011rs}. We develop a novel EFT for the \X\ with non-relativistic fields for $D^0,\,\bar{D}^0$ and $\pi^0$, demanding exact Galilean invariance. The theory allows for systematic calculations of both the pole and the line shape of a bound \X\ state in $D^0\bar{D}^0\pi^0$ production, including theoretical uncertainties. Thereby, it provides a tool for the extraction of the mass and width of the \X\ from an experimental peak that is influenced by threshold effects. Our power counting infers the perturbativeness of pions and charged mesons from the characteristic momentum scales of the system. Moreover, it suggests that some ingredients in the approach of Ref.~\cite{Baru:2011rs} can be neglected at NLO. This makes the theory renormalizable for arbitrary cutoffs.

The paper is organized as follows. In Sec.~\ref{Sec:EFTLagrangian}, we construct the Galilean-invariant EFT Lagrangian and introduce the $D^{0\ast}$ ($\bar{D}^{0\ast}$) as a $p$-wave resonance in the $D^0\pi^0$ ($\bar{D}^0\pi^0$) sector. Based on a comprehensive scaling analysis of $D^0\pi^0$ threshold parameters, we determine an appropriate $D^{0\ast}$ propagator expansion in Sec.~\ref{Sec:DPi}. Moreover, we effectively include radiative decays of the $D^{0\ast}$ using complex self-interactions. In Sec.~\ref{Sec:DDbarPi}, we construct the non-perturbative $D^0\bar{D}^{0\ast}$ amplitude and solve it analytically at leading order (LO). Afterward, we show that $D^{0\ast}$ self-interactions, pion exchanges and charged mesons enter at NLO. Finally, Sec.~\ref{Sec:Results} presents numerical results for $\GX$ and the line shape at LO and NLO. Moreover, it is illustrated how the \X\ peak gets modified by the detector's energy resolution. We conclude with a summary and an outlook in Sec.~\ref{Sec:SummaryOutlook}.

\section{EFT Lagrangian}
\label{Sec:EFTLagrangian}

We start by constructing a Galilean-invariant Lagrangian $\mathcal{L}$ for particles $D^0$, $\bar{D}^0$ and $\pi^0$. For clarity, we decompose $\mathcal{L}$ into different scattering sectors by writing
\begin{equation}
	\mathcal{L} = \mathcal{L}_\text{kin}
	+ (\mathcal{L}_{D\pi} +\mathcal{L}_{\bar{D}\pi})
	+\mathcal{L}_{D\bar{D}\pi}\,.
\end{equation}
The kinetic part of the EFT Lagrangian,
\begin{equation}
	\label{Eq:LKin}
	\mathcal{L}_\text{kin}= 
	D^\dagger
	\left[
		i\,\partial_0+\frac{\nabla^2}{2m_D}
	\right]
	D
	+ \bar{D}^\dagger
	\left[
		i\,\partial_0+\frac{\nabla^2}{2m_D}
	\right]
	\bar{D}
	+ \pi^\dagger
	\left[
		i\,\partial_0+\frac{\nabla^2}{2m_{\pi}}
	\right]
	\pi\,,
\end{equation}
contains fields with masses \mbox{$m_D\equiv \unit[1864.83(5)]{MeV}$} and \mbox{$m_\pi\equiv \unit[134.9770(5)]{MeV}$} determined from experiment \cite{PDG2017}. All rest masses have been shifted to zero. Note, that also charged mesons will enter the EFT at NLO. Their inclusion will be discussed in Sec.~\ref{Sec:DDbarPi}.

Similar to neutron-alpha scattering, the resonance $D^{0\ast}$ ($\bar{D}^{0\ast}$) can be treated by an auxiliary field \cite{Bedaque:2003wa}. Accordingly, we introduce the vector field $\bs{D}$ ($\bs{\bar{D}}$) in $\mathcal{L}_{D\pi}$ ($\mathcal{L}_{\bar{D}\pi}$). It encapsulates all $p$-wave interactions and could in principle be eliminated by performing the Gaussian path integral over $\bs{D}$ ($\bs{\bar{D}}$) or using the equations of motion. We write
\begin{align}	\label{Eq:LDPi}
	\mathcal{L}_{D\pi} =\
	&\bs{D}^\dagger
	\left[
		\Delta_0+\Delta_1\,i\,\partial_\text{cm}
		+\sum_{n\geq 2}\Delta_n\,(i\,\partial_\text{cm})^n
	\right] 
	\bs{D}
	+ g \left[
		\bs{D}^\dagger \cdot \big(
			\pi\overleftrightarrow{\bs{\nabla}}D
		\big)
		+ \text{H.c.}
	\right],
\end{align}
where ``H.c.'' denotes the Hermitian conjugate. The first term of Eq.~\eqref{Eq:LDPi} defines the bare $D^{0\ast}$ propagator. It is given by a series in the Galilean-invariant derivative \mbox{$i\,\partial_\text{cm} \equiv i\,\partial_0 +\nabla^2/(2M)$} with total $D^0\pi^0$ mass $M\equiv m_D+m_\pi$. 
This form ensures analyticity in the center-of-mass energy $E_\text{cm}\equiv
	E-\bs{p}_{D^{0\ast}}^2/(2M)$ and reproduces the $D^0\pi^0$ effective range expansion in Sec.~\ref{Sec:DPi}.
The real-valued coefficients $\Delta_{n\geq 0}$ ($\Delta_1\equiv \pm 1$) have mass units $\unit{MeV}^{1-n}$. Note, that the sign $\Delta_1$ cannot be changed by field redefinitions and has to be determined in the re\-nor\-ma\-li\-za\-ti\-on procedure. In our case, $\Delta_1= +1$ (see Appendix~\ref{Sec:Appendix2Body} for details). Therefore, $\bs{D}^{(\dagger)}$ is a physical field. Later, radiative decays $D^{0\ast}\rightarrow D^0\gamma$ ($\bar{D}^{0\ast}\rightarrow \bar{D}^0\gamma$) will by included by adding imaginary parts to the $\Delta_n$.

\begin{figure}
	\begin{center}
	\subfigure[]{
		\includegraphics[scale=1]{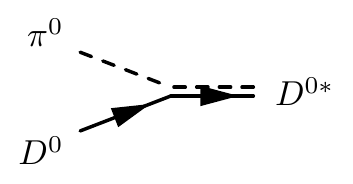}
	}
	\subfigure[]{
		\includegraphics[scale=1]{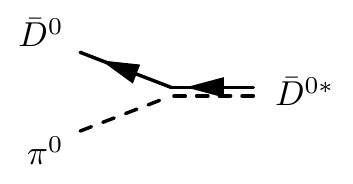}
	}
	\subfigure[]{
		\includegraphics[scale=1]{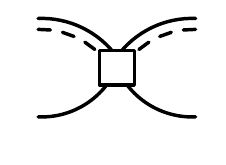}
	}
	\vspace{-0.2cm}
	\caption{\label{Fig:VertexDiagrams}Vertex diagrams. The Feynman
		rules for (a) $D^0\pi^0\rightarrow D^{0\ast}$
		and (b) $\bar{D}^0\pi^0\rightarrow \bar{D}^{0\ast}$
		read $-g\,k_j$ with relative incoming momentum $\bs{k}$ and
		vector meson polarization $j\in\{1,\,2,\,3\}$.
		Vertex (c) reads $-i\,C_0\,\delta^{ij}$ and connects $C=+$
		states $(D^0\bar{D}^{0\ast}+\bar{D}^0 D^{0\ast})/\sqrt{2}$.
		Thus, flavor-indicating arrows are omitted.
	}
	\end{center}
	\vspace{-0.5cm}
\end{figure}

The second term of Eq.~\eqref{Eq:LDPi} allows for transitions $D^0\pi^0\leftrightarrow D^{0\ast}$ and is depicted in Fig.~\ref{Fig:VertexDiagrams}(a). It depends on the coupling constant $g$ and the Galilean-invariant derivative
\mbox{$\overleftrightarrow{\bs{\nabla}} \equiv\ \mu\,\big(
	m_\pi^{-1}\,\overleftarrow{\bs{\nabla}}
	-m_D^{-1}\,\overrightarrow{\bs{\nabla}}
\big)$}
with reduced $D^0\pi^0$ mass $\mu\equiv (m_D^{-1} + m_\pi^{-1})^{-1}$.
Thus, Feynman rules for these transitions depend on the relative $D^0\pi^0$ momentum.
The values of all parameters in $\mathcal{L}_{D\pi}$ will be addressed in Sec.~\ref{Sec:DPi}. As a consequence of the charge-conjugation symmetry of the \X, the Lagrangian part $\mathcal{L}_{\bar{D}\pi}$ is obtained by replacing \mbox{$(D,\,\bs{D}) \rightarrow (\bar{D},\,\bs{\bar{D}})$} in Eq.~\eqref{Eq:LDPi}.

The three-body part of the Lagrangian,
\begin{equation}
	\label{Eq:Contact}
	\mathcal{L}_{D\bar{D}\pi}
	=\ -C_0\,\frac{1}{2}\left[
		\bar{D}\bs{D} + D \bs{\bar{D}}
	\right]^\dagger \cdot
	\left[
		\bar{D}\bs{D} + D \bs{\bar{D}}
	\right]
        +\ldots,
\end{equation}
contains all $D^0\bar{D}^{0\ast}$\ $s$-wave interactions in the $C=+$ channel. In Sec.~\ref{Sec:DDbarPi}, the coupling $C_0$ will be used to generate the \X. Its vertex is depicted in Fig.~\ref{Fig:VertexDiagrams}(c).
Higher-order terms contained in the ellipsis
do not enter up to NLO (see Sec.~\ref{Sec:DDbarPi}).

Throughout the paper we label relative $D^0\pi^0$ momenta with $k$ or $l$ and relative $D^0\bar{D}^{0\ast}$ momenta with $p$ or $q$. Moreover, we drop the particle superscripts $0$ from now on.

\section{Two-Body Sector: The {$\bs{D^{0\ast}}$} Resonance}
\label{Sec:DPi}

The \X\ lies extremely close to the $\bar{D}D^\ast$ ($D\bar{D}^\ast$) threshold. Therefore, its form crucially depends on the vector meson propagator. It is the goal of this section to identify an appropriate propagator expansion in the vicinity of the \X. Without loss of generality we focus on the $D^\ast$ system. We recover the $D\pi$ effective range expansion from the EFT Lagrangian and analyze it in terms of characteristic momentum scales. Thereby, we obtain a natural explanation for the narrowness of the $D^\ast$ resonance. Afterward, we include radiative decays of the $D^\ast$. The resulting expansion of the propagator is given at the end of the section.

\subsection{Matching to the Effective Range Expansion}

All terms in Eq.~\eqref{Eq:LDPi} are Galilean-invariant and potentially contribute to the $D^\ast$ propagator. However, in order to produce the $D^\ast$ as a $p$-wave resonance, only a few terms are needed. This statement will be verified in this section. We begin by matching the propagator terms to the effective range expansion of the $D\pi$ amplitude.

\begin{figure}
	\begin{center}
		\includegraphics[scale=1]{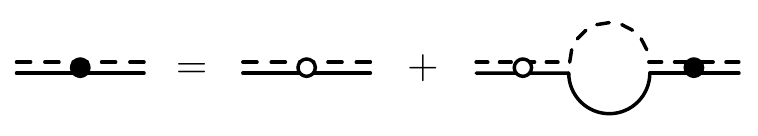}
		\vspace{-0.5cm}
	\caption{\label{Fig:FeynDyson}The full $D^\ast$ ($\bar{D}^\ast$) propagator (indicated by a black dot) contains iterations of the bare one (indicated by an empty dot) and the self-energy loop $-i\,\Sigma$. Flavor-indicating arrows are omitted.}
	\end{center}
\end{figure}

Let $p^\mu=(p^0,\bs{p})$ be the $D^\ast$ four-momentum. The bare propagator connecting equal polarization states at a center-of-mass energy $\Ecm= p^0-\bs{p}^2/(2M)$ is given by
\begin{equation}	\label{Eq:GBare}
	i\,G_\ast^\text{(b)}(E_\text{cm})=i
	\left[
		\Delta_0 + (E_\text{cm}+i\epsilon)
		+\sum_{n\geq 2}\Delta_n\,(E_\text{cm}+i\epsilon)^n
	\right]^{-1},
\end{equation}
where we have used $\Delta_1=+1$. To obtain the full propagator, the bare one needs to be dressed by $D^\ast$ self-energy loops $-i\,\Sigma(\Ecm)$ as shown in Fig.~\ref{Fig:FeynDyson}. We then obtain
\begin{equation}	\label{Eq:GFull}
	i\,G_\ast^\text{(f)}(E_\text{cm})=i\,\left[
		{G_{\ast}^\text{(b)}}^{-1}(\Ecm)-\Sigma(\Ecm)
	\right]^{-1}.
\end{equation}

In Appendix~\ref{Sec:Appendix2Body}, we calculate $\Sigma$ using dimensional regularization and the Power Divergence Subtraction Scheme (PDS) with renormalization scale $\Lambda_\text{PDS}$ \cite{Kaplan:1998tg}. For convenience, we 
use Minimal Subtraction (MS) instead of PDS for all practical calculations
by considering the limit $\Lambda_\text{PDS}\rightarrow 0$. This choice makes our scaling analysis much more transparent since all quantities are automatically
scale-independent in the MS scheme. In any case, observables do not depend on the chosen renormalization scheme. For the self energy, we obtain
\begin{equation}	\label{Eq:Sigma}
	\Sigma(E_\text{cm})= -g^2 \frac{\mu}{6\pi}
	\left[-2\mu (E_\text{cm}+i\epsilon)\right]^{3/2},
\end{equation}
which is purely imaginary for $\Ecm>0$.

In the auxiliary field formalism, one gets the elastic $D\pi$ scattering amplitude $t$ for relative momenta $\bs{k},\,\bs{k}'$ by attaching external $D\pi$ legs to the full $D^\ast$ propagator at $\Ecm=k^2/(2\mu)$, i.e., \mbox{$t(\bs{k},\,\bs{k}')=
	-g^2\,\bs{k}\cdot\bs{k}'\, G_\ast^\text{(f)}\big(k^2/(2\mu)\big)$}. This expression can be matched to the generic $p$-wave effective range expansion \cite{Bertulani:2002sz},
\begin{equation}	\label{Eq:tOS}
	t(\bs{k},\,\bs{k}') = \frac{6\pi}{\mu}\,\bs{k}\cdot\bs{k}'\left[
		-a_1^{-1} + \frac{r_1}{2}\, k^2 - i\, k^3
		+\sum_{n\geq 2}
		\mathcal{P}_{2n}\,k^{2n}
	\right]^{-1}.
\end{equation}
The parameters $a_1$ (``scattering volume'') and $r_1$ (``$p$-wave effective range'') have mass dimensions $\unit{MeV^{-3}}$ and $\unit{MeV}$, respectively. The coefficients $\mathcal{P}_{2n}$ will be referred to as ``higher-order parameters'' from now on. Comparing Eq.~\eqref{Eq:GFull} and \eqref{Eq:tOS}, we identify
\begin{equation}	\label{Eq:Identities}
	-a_1^{-1}=-\frac{6\pi}{\mu}\,\frac{\Delta_0}{g^2}\,,
	\qquad
	\,\frac{r_1}{2}=-\frac{6\pi}{\mu}\,\frac{1
	}{2\mu g^2}\,,
	\qquad
	\mathcal{P}_{2n}
	=-\frac{6\pi}{\mu}\,\frac{\Delta_n}{(2\mu)^n\,g^2}
	\quad (n\geq 2)\,.
\end{equation}
We are now in the position to analyze the  characteristic momentum scales
of the propagator.

\subsection{Momentum Scales and Scaling Analysis at Threshold}

The $D^\ast$ resonance occurs at a $D\pi$ center-of-mass energy $\delta = \unit[7.04(3)]{MeV}$, see Eq.~\eqref{Eq:Delta}, which is small compared to all involved particle masses \cite{PDG2017}. This observation gives rise to a significant separation of momentum scales, which can be explained by a fine-tuning of the underlying theory, QCD. The relative $D\pi$ momentum $k$ needed to probe the shallow resonance is of the order \mbox{$k\sim \KLO\equiv (2\mu\,\delta)^{1/2}\approx \unit[42]{MeV}$}. In contrast, the natural momentum scale $\KHI\sim m_\pi$ occurring in QCD is much larger. Due to the fact that our EFT is non-relativistic, $\KHI$ represents its breakdown point. In the following, we express the scale separation as the ratio
\begin{equation}
	\kappa\equiv \KLO/\KHI\sim 0.3\,.
\end{equation}
Note that the effective range expansion is in $k^2$. As a consequence, the $D^\ast$ propagator expansion will be in $\kappa^2\sim 0.09$. Thus, we expect quick convergence.

\subsubsection{Naturalness of Higher-Order Parameters}

Each threshold parameter in Eq.~\eqref{Eq:tOS} scales with certain powers of $\KLO$ and $\KHI$. We assume that fine-tunings related to the shallow $D^\ast$ resonance occur in $a_1$, $r_1$ or in both of them. In contrast, all higher-order parameters are assumed to be of natural size, i.e., $\mathcal{P}_{2n}\sim \KHI^{3-2n}$. This ``naturalness argument'' is based on the assumption that a scaling scenario with as few fine-tunings as possible is most likely to occur in nature \cite{tHooft:1979rat,Bedaque:2003wa}.

As a consequence, the terms \mbox{$\mathcal{P}_{2n}k^{2n}\sim \kappa^{2n}\KHI^{3}$} in Eq.~\eqref{Eq:tOS} are suppressed compared to the unitary cut $|i\,k^3|\sim \KLO^3$ by at least one order in $\kappa$. In other terms, due to Eq.~\eqref{Eq:Identities}, higher-order propagator terms close to resonance are suppressed compared to the self-energy like
\begin{equation}
	\label{Eq:DeltanSigmaSuppression}
	\Delta_n\, \delta^n \sim \kappa^{-3+2n}\,|\Sigma(\delta)|
	\quad (n\geq 2)\,.
\end{equation}

\subsubsection{Consequences from the Small $D^{0\ast}$ Width}

We have seen that natural $\mathcal{P}_{2n}$ imply strongly suppressed higher-order propagator terms. With this information at hand, we can now express $\Delta_0$ and the coupling $g$ in terms of the $D^\ast$ resonance parameters. The $D^\ast$ resonance manifests itself as a complex energy pole in $i\,G_\ast^\text{(f)}(\Ecm)$. It can be parametrized by $\delta-i\,\GDPi/2$, where $\GDPi$ denotes the small width of the decay $D^\ast\rightarrow D\pi$. Although $\GDPi$ has not been measured yet, an upper bound\footnote{This number already includes the branching ratio of the pionic decay channel.} $\GDPi<\unit[1.4]{MeV}$ is known from experiment~\cite{PDG2017}. Thus, the ratio $\chi\equiv\GDPi/(2\delta)<0.1$ is very small.
Demanding $0\equiv\ {G_\ast^\text{(f)}}^{-1}(\delta-i\,\GDPi/2)$ and using Eq.~\eqref{Eq:DeltanSigmaSuppression}, we obtain
\begin{equation}
	\label{Eq:GammaGCorrelation}
	-\Delta_0=\ 
	\delta\,\Big(
		1+\mathcal{O}\left(\chi\,\kappa\right)
	\Big)\,,
	\qquad
	|\Sigma(\delta)|=g^2\,\frac{\mu}{6\pi}(2\mu\,\delta)^{3/2}
	=\frac{\GDPi}{2}\,\Big(
		1+\mathcal{O}\left(\chi\,\kappa\right)
	\Big)
\end{equation}
with $\kappa\sim 0.3$. We see that the width $\GDPi$ is given by $2|\Sigma(\delta)|$, up to a tiny uncertainty of order $\chi\,\kappa < \unit[3]{\%}$. This estimation stems from the product $\chi\,\mathcal{P}_4\,\KLO$ in the case of $\mathcal{P}_4\sim \KHI^{-1}$ being natural. Thus, for the approximations in Eq.~\eqref{Eq:GammaGCorrelation} to fail, $\mathcal{P}_4$ would have to be enhanced by a factor of order $(\chi\,\kappa)^{-1} > 33$, which is unlikely. In fact, we will find below that $\chi\ll 0.1$, which secures the validity of the above approximations.

\subsubsection{Parameter Fixing}

We follow Braaten~\cite{Braaten:2015tga}
and infer a value for $g^2$ (and thus of $\GDPi$) from the total pionic
decay width $\Gc\equiv\Gamma[D^{+\,\ast}\rightarrow D^0\pi^+ + D^+\pi^0]
=\unit[82(2)]{keV}$ 
of the charged $D^{+\,\ast}$ meson using a modified version of
Eq.~\eqref{Eq:GammaGCorrelation} and isospin symmetry
(see Appendix~\ref{Sec:AppendixCoupling}
        for details). This yields
\begin{equation}
	\label{Eq:GAndGDPi}
	g^2= \unit[3.48(8)\cdot 10^{-8}]{MeV^{-3}}\,,
	\qquad
	\GDPi= \unit[34.7(9)]{keV}\,.
\end{equation}
The indicated uncertainties are of order $\unit[3]{\%}$ and arise from the experimental uncertainty of $\Gc$ only. In contrast, uncertainties from natural higher-order parameters are negligible. As indicated above, the actual width $\GDPi$ is indeed much smaller than the experimental bound. It implies a tiny ratio $\GDPi/(2\delta)=\chi\approx 0.0025\ll 0.1$ and thus, in the case of natural $\mathcal{P}_{2n}$, a theoretical uncertainty of order $\chi\,\kappa\sim \unit[0.075]{\%}\ll \unit[3]{\%}$. For this reason, we use the central values of $g^2$ and $\GDPi$ of Eq.~\eqref{Eq:GAndGDPi} in all later calculations.

Using the value of the coupling, we can now calculate $a_1$ and $r_1/2$ from Eq.~\eqref{Eq:Identities} at the experimental uncertainty level of $\Gc$, yielding
\begin{equation}
	\label{Eq:NumericResultsDPi}
	a_1^{-1/3}=-\unit[312(3)]{MeV}\,,
	\quad
	\frac{r_1}{2}=-\unit[17.1(4)]{GeV}\,.
\end{equation}

\subsubsection{Scaling of $a_1$ and $r_1/2$ and Fine-Tuning Scenarios}

Given the numerical values of Eq.~\eqref{Eq:NumericResultsDPi} we are now able to assess scaling situations for $a_1$ and $r_1/2$ that have been used in the literature for other physical systems. First of all, we see from Eq.~\eqref{Eq:Identities} and \eqref{Eq:GammaGCorrelation} that the inverse scattering volume and the unitary cut are separated like $|i\,\KLO^3/a_1^{-1}|=|\Sigma(\delta)/\Delta_0|\approx \GDPi/(2\delta)=\chi$. Moreover, we know that $|i\,\KLO^3|\gg \mathcal{P}_{2n}\KLO^{2n}$. It follows that the resonance pole can only occur if  $a_1^{-1}\sim r_1/2\,\KLO^2$.

One scenario that respects this pole condition has been discussed by Bertulani~\etal\ for neutron-alpha scattering. They analyzed the situation, in which both $a_1^{-1}$ and $r_1/2$ are unnaturally small, i.e., $a_1^{-1}\sim \KLO^3$ and $r_1/2\sim\KLO$ \cite{Bertulani:2002sz}. Due to Eq.~\eqref{Eq:Identities} this scheme requires two fine-tuned combinations of coupling constants, $\Delta_0/g^2$ and $1/g^2$. Bedaque~\etal\ have argued that such a high degree of fine-tuning is unlikely to occur in nature. Their modified scheme $a_1^{-1}\sim \KLO^2\KHI$ and $r_1/2\sim \KHI$ only requires $\Delta_0/g^2$ to be unnaturally small \cite{Bedaque:2003wa}. Yet, in the case of $D\pi$ scattering both schemes appear to be inappropriate since the value of $|r_1/2|$ in Eq.~\eqref{Eq:NumericResultsDPi} exceeds $\KHI$ by several orders of magnitude.

For the $D\pi$ sector, we propose the novel scheme
\begin{equation}
	a_1^{-1}\sim \KHI^3\,,\qquad \frac{r_1}{2}\sim\KLO^{-2}\KHI^3\,,
\end{equation}
in which only the $p$-wave effective range is enhanced. This explains its huge numerical value and also the rather natural value $|a_1^{-1/3}|\approx 2.3\, \KHI$. Further evidence for this scheme comes from the fact that only one combination of constants, \mbox{$1/(2\mu g^2)\sim \kappa^{-2}\KHI\, \mu/(6\pi)$}, needs to be fine tuned, while the combination \mbox{$\Delta_0/g^2\sim \KHI^3\,\mu/(6\pi)$} scales naturally. Note that there are other scaling scenarios consistent with the pole condition that could explain the specific values of $a_1^{-1}$ and $r_1/2$. Yet, they would inevitably involve two or more fine-tunings.

In our scheme, the width $\GDPi/2\approx |\Sigma(\delta)|$ is suppressed by three orders with respect to $\delta$, i.e., \mbox{$\chi=\GDPi/(2\delta)\sim \kappa^3$}. This shows that the narrowness of the $D^\ast$ resonance can be explained naturally by the momentum scales of the $D\pi$ system.

\subsection{Extension for Radiative $\bs{D^{0\ast}}$ Decays}

The radiative decay $D^\ast\rightarrow D\gamma$ has a large branching ratio $\mathcal{B}\equiv \GDg/(\GDPi+\GDg)=\unit[35.3(9)]{\%}$ \cite{PDG2017}. It translates to the width $\GDg=\unit[18.9(9)]{keV}\approx 0.55\,\GDPi$. Thus, it is as important for the \X\ resonance, and we count $\GDg\sim\GDPi$. The full $D^\ast$ pole position reads
\begin{equation}\label{Eq:GammaStarFullError}
	E_\ast\equiv \delta-i\,\frac{\Gast}{2}\,,
	\qquad
	\Gast\equiv \GDPi+\GDg= \unit[(53.6\pm 1.0)]{keV}\,.
\end{equation}
In Sec.~\ref{Sec:DDbarPi} we argue that the width $\GX$ at LO is just given by $\Gast$. In fact, numerical results will suggest that this approximation is even valid up to NLO.

The relative momentum of the decay products $D\gamma$ is given by $\unit[137]{MeV}\sim \KHI$, which lies beyond the scope of our EFT \cite{PDG2017}. Still, $\GDg$ can be introduced effectively by adding an anti-Hermitian part to the Lagrangian \cite{Braaten:2016sja}, i.e., we replace \mbox{$\Delta_n\rightarrow \Delta_n + i\,W_n$}. Note, that the former relations
\begin{equation}
	\label{Eq:PropTermRelReal}
	\Delta_0\approx -\delta\,,
	\qquad
	\Delta_1\equiv 1\,,
	\qquad
	\Delta_n\sim \chi\,\kappa^{-3+2n}\,\delta^{1-n}
	\quad(n\geq 2)
\end{equation}
shall not be affected by this procedure. Due to $\GDg/2\sim \chi\, \delta\sim \chi\,\kappa^2\,\KHI^2/(2\mu)$, all imaginary parts $W_{n\geq 0}$ must involve a common suppression factor $\chi\,\kappa^2$. There is no reason to assume fine-tunings between different $W_n$. Thus, we count
\begin{equation}
	\label{Eq:PropTermRelImag}
	W_n\sim \chi\,\kappa^2 \left(\frac{\KHI^2}{2\mu}\right)^{1-n}\sim
	\chi\,\kappa^{2n}\,\delta^{1-n}
	\quad (n\geq 0)\,.
\end{equation}
Demanding $0\ \equiv\ {G_\ast^\text{(f)}}^{-1}(E_\ast)$, we recover Eq.~\eqref{Eq:GammaGCorrelation} and find $W_0=\GDg/2\,\big(1+\mathcal{O}(\kappa^2)\big)$.

\subsection{Propagator Expansion at Resonance}

Our findings show that close to the $D\pi$ threshold, the self-energy and all propagator terms proportional to $\Delta_{n\geq 2}$ or $W_{n\geq 0}$ are suppressed compared to $\delta$. Thus, the width is a sub-leading phenomenon. However, the \X\ resonance does not occur at the $D\pi$ threshold but in the immediate vicinity of the $D^\ast$ resonance, i.e., at $\Ecm\approx E_\ast$. In this region, $\Delta_0$ and $\operatorname{Re}\Ecm$ almost cancel, resulting in a much weaker suppression of the $D^\ast$ width. As discussed by Bedaque \etal\ \cite{Bedaque:2003wa}, an appropriate ordering scheme must take into account this kinematic fine-tuning $|\Ecm - E_\ast|/|E_\ast| \ll 1$. As a consequence, close to resonance, we have to include the width nonperturbatively.

\begin{figure}
	\includegraphics[scale=1]{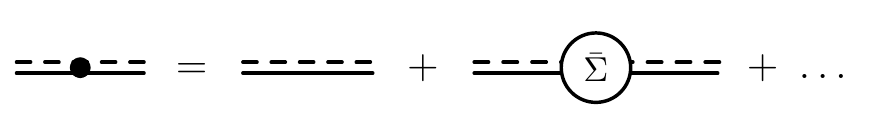}
	\vspace{-0.3cm}
	\caption{\label{Fig:PropCorrs}
		$D^\ast$ propagator expansion at resonance. The first correction to the LO propagator $i\,G_\ast$ (without any dot) contains the self-energy insertion $-i\,\bar{\Sigma}$. Flavor-indicating arrows are omitted.
	}
\end{figure}

Again, we consider the full $D^\ast$ propagator of Eq.~\eqref{Eq:GFull} including radiative decays. The pole position $E_\ast$ becomes most apparent in the form
\begin{align}
	\label{Eq:GFullComplex}
	i\,G_\ast^\text{(f)}(\Ecm)
	=\ i\,\Bigg[
		\Big(1+i\,W_1\Big)\,\Ecm
		- \Sigma(\Ecm)
		+\sum_{n\geq 2}\left(\Delta_n+i\,W_n\right)
		\Ecm^n
		-\left[\Ecm\rightarrow E_\ast\right]
	\Bigg]^{-1}.
\end{align}
Note, that the expression $[\Ecm\rightarrow E_\ast]$ is just given by the constant $-(\Delta_0+i\,W_0)$. We make the kinematic fine-tuning explicit by factoring out the term
\begin{equation}
	\label{Eq:GStarLO}
	i\,G_\ast(\Ecm)\equiv \frac{i}{\Ecm-E_\ast+i\,\epsilon}\,,
\end{equation}
which will be the leading-order $D^\ast$ propagator for the calculation of the \X\ width. It exhibits a Breit-Wigner form and is depicted by a simple straight-dashed double line; see Fig.~\ref{Fig:PropCorrs}.

After the factorization, all terms in the brackets of Eq.~\eqref{Eq:GFullComplex} -- besides the leading $1$ -- are at least suppressed by the factor $\chi=\GDPi/(2\delta)\ll 1$ [see Eq.~\eqref{Eq:GammaGCorrelation}, \eqref{Eq:PropTermRelReal} and \eqref{Eq:PropTermRelImag}] and thus very small. Therefore, we refer to them as ``propagator corrections''. The most important one involves the self-energy and a counter term $-i\,\Sigma(E_\ast)$. It reads
\begin{equation}
	\label{Eq:PropCorrsAtRes1}
	-i\,\bar{\Sigma}(\Ecm)\cdot i\,G_\ast(\Ecm)
	\equiv\frac{\Sigma(\Ecm)-\Sigma(E_\ast)}{\Ecm-E_\ast}
	\sim\ \Sigma'(E_\ast)
	=\mathcal{O}\left(\chi^2\right)
	+i\,\mathcal{O}\left(\chi\right).
\end{equation}
The indicated scalings for the real and imaginary part follow from $E_\ast=\delta\,\big(1+i\,\mathcal{O}(\chi)\big)$ and $\Sigma'(E_\ast)\sim \Sigma(E_\ast)/E_\ast$. Note, that they hold not only in the region $\Ecm\approx E_\ast$ but also at $\Ecm\approx 0$.\footnote{For $\Ecm=0$, the difference quotient in Eq.~\eqref{Eq:PropCorrsAtRes1} collapses to $\Sigma(E_\ast)/E_\ast$.} In fact, all propagator corrections exhibit this feature. It is crucial for this work because \textit{both} regions are important for the \X\ width power counting in Sec.~\ref{Sec:DDbarPi}. The propagator expansion is depicted in Fig.~\ref{Fig:PropCorrs} and reads
\begin{align}
	\label{Eq:GFullComplexFactorized}
	i\,G_\ast^\text{(f)}(\Ecm)
	=&\ 
	i\,G_\ast(\Ecm)\ 
	\Bigg[\,
		1
		-i\,\bar{\Sigma}(\Ecm)\cdot i\,G_\ast(\Ecm)
		+ \mathcal{O}\left(\chi\,\kappa\right)
		+i\,\mathcal{O}\left(\chi\,\kappa^2\right)
	\,\Bigg]\,.
\end{align}
All propagator corrections involving multiple self-energies or any of the coefficients $\Delta_{n\geq 2},\,i\,W_{n\geq 1}$ are condensed into the expression $\mathcal{O}(\chi\,\kappa)+i\,\mathcal{O}(\chi\,\kappa^2)$.

As we will see in Sec.~\ref{Sec:DDbarPi}, modifications to the \X\ width are determined by the imaginary parts of the propagator corrections in Eq.~\eqref{Eq:GFullComplexFactorized}. Thus, the self-energy correction with its imaginary part $\sim \chi\,\kappa^0$ is the most important one. All imaginary parts exhibit even powers in $\kappa$ ($\chi\,\kappa^0,\,\chi\,\kappa^2,\,\chi\,\kappa^4,$ etc.). This observation confirms that the expansion is indeed in $\kappa^2\sim 0.09$.

There is one more reason why the given expansion is beneficial for our purposes: for a finite number of corrections there is always only one energy pole representing the $D^\ast$. Therefore, we do not have to take into account additional unphysical $D\pi$ states \cite{Bertulani:2002sz}. Finally, let us mention that relativistic corrections in the $D^\ast$ propagator can be neglected at the order we work. For more details, we refer to Appendix~\ref{Sec:AppendixRelativistics}.

\section{Three-Body Sector: The {$\bs{X(3872)}$}}
\label{Sec:DDbarPi}

We now use the $D^\ast$ propagator expansion to produce the \X\ as an energy pole in the $D\bar{D}^\ast$ amplitude $T$. First, we diagrammatically construct the amplitude in the $J^{PC}=1^{++}$ channel and explain the renormalization procedure. Moreover, we show how $T$ can be used to calculate the line shape of the \X\ in $D\bar{D}\pi$ production.

Afterward, we analyze all considered $D\bar{D}^\ast$ interactions according to their influence on the \X\ width. It turns out that corrections to the LO width $\GX^\text{(LO)}=\Gast$ are suppressed by factors comparable to $\kappa^2\sim 0.09$. We identify all NLO corrections in a diagrammatic power counting, which exploits the characteristic momentum scales of the system.

\subsection{Non-perturbative $\bs{D^0\bar{D}^{0\ast}}$ Amplitude}

Similarly to the two-body sector, we start by constructing the $D\bar{D}^\ast\rightarrow D\bar{D}^\ast$ transition amplitude $T$ nonperturbatively. Note, that the unstable $D^\ast$ has a complex rest mass $E_\ast$. Thus, it can never occur as an asymptotic state. Still, the amplitude $T$ is needed to connect intermediate $D\bar{D}^\ast$ states, e.g. in $D\bar{D}\pi$ production (see Fig.~\ref{Fig:DDbarPiProd}).

\subsubsection{Diagrammatic Construction}

From $\mathcal{L}_{D\pi}$ and $\mathcal{L}_{D\bar{D}\pi}$ in Eqs.~\eqref{Eq:LDPi} and \eqref{Eq:Contact} we see that the particles $D\bar{D}^\ast$ can either exchange a pion or interact through the contact interaction $C_0$.
Moreover, we allow for an additional $s$-wave $D\bar{D}^\ast$ interaction $-i\,\mathcal{I}_\text{c}$ that takes into account NLO contributions from charged states $(D^+D^{-\ast}+D^-D^{+\ast})/\sqrt{2}$. Its analytic form is given in Eq.~\eqref{Eq:ChargedInteraction} below. In Sec.~\ref{Sec:DDbarPi} D, we show explicitly that only $s$-wave contact interactions enter $-i\,\mathcal{I}_\text{c}$ at NLO. Thus, it acts only in the $L=0$ channel and does not change the vector meson polarization.

\begin{figure}
	\begin{center}
		\includegraphics[scale=1]{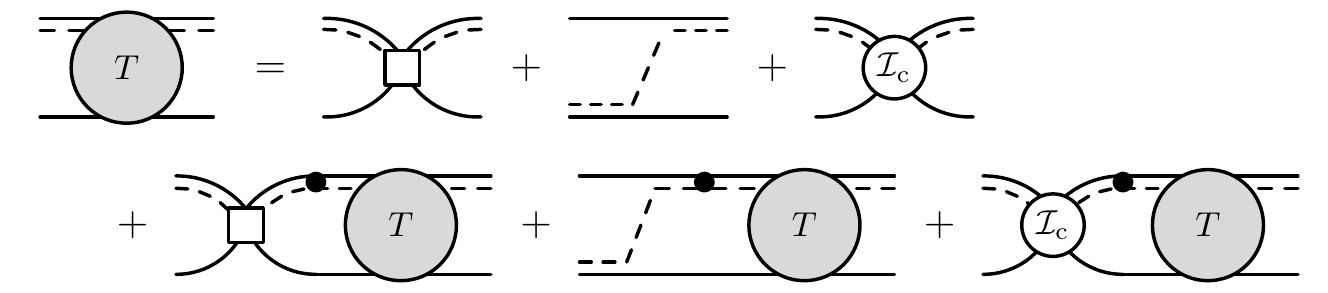}
	\end{center}
	\vspace{-0.5cm}
	\caption{\label{Fig:3BodyAmp}Non-perturbative $D\bar{D}^\ast$ amplitude $T$ in the $C=+$ channel.}
\end{figure}

By iterating all these interactions in the $C=+$ channel, we obtain an integral equation for the amplitude $T$ which is shown diagrammatically in Fig.~\ref{Fig:3BodyAmp}. Flavor-indicating arrows are left out since each $D\bar{D}^\ast$ state is in a $C=+$ superposition. Every loop is associated with an integral over the intermediate relative four-momentum \mbox{$q^\mu\equiv (q^0,\,\bs{q})\equiv \mu_\ast\left(m_D^{-1}\,q^\mu_D-M^{-1}\,q^\mu_{\bar{D}^\ast}\right)$} with relative $D\bar{D}^\ast$ mass $\mu_\ast\equiv (m_D^{-1}+M^{-1})^{-1}$. The choice of this integration variable is not mandatory but convenient as it exploits Galilean symmetry.

We investigate the process in the center-of-mass frame, i.e., in the incoming (outgoing) channel, we set $\bs{p}\equiv\bs{p}_D=-\bs{p}_{\bar{D}^\ast}$ ($\bs{p}'\equiv\bs{p}_D'=-\bs{p}_{\bar{D}^\ast}'$). The energy $E$ is defined relative to the $D\bar{D}\pi$ threshold. The $q^0$ integration can be performed with the residue theorem, leading to
\begin{align}
	\label{Eq:3BodyAmp}
	\nonumber
	i\,T^{\,ij}\left(\bs{p},\,\bs{p}';\,E\right)
	=\ &	-i\,C_0\,\delta^{ij}
	-i\,V^{ij}\left(\bs{p},\,\bs{p}';\,E\right)
	-i\,\mathcal{I}_\text{c}(E)\,\delta^{ij}
	\\
	&+\sum_{r=1}^3 \int\!\frac{\text{d}^3q}{(2\pi)^3}\,
	\frac{
		C_0\,\delta^{ir}
		+V^{ir}\left(\bs{p},\,\bs{q};\,E\right)
		+\mathcal{I}_\text{c}(E)\,\delta^{ir}
	}{
		{G_\ast^\text{(f)}}^{-1}\Big(E-\bs{q}^2/(2\mu_\ast)\Big)
	}\,i\,T^{\,rj}\left(\bs{q},\,\bs{p}';\,E\right)
\end{align}
for $D^\ast$ polarizations $i$ and $j$. The pion exchange potential is given by
\begin{equation}
	\label{Eq:PionExPot}
	i\,V^{ij}\left(\bs{p},\,\bs{q};\,E\right)
	\equiv i\,g^2
	\frac{
		\left(\alpha\,\bs{p}+\bs{q}\right)^i
		\left(\alpha\,\bs{q}+\bs{p}\right)^j
	}{
		E
		-\frac{\bs{p}^2}{2\mu}
		-\frac{\bs{q}^2}{2\mu}
		-\frac{\bs{p}\cdot\bs{q}}{m_\pi}
		+i\epsilon
	}\,,
\end{equation}
where we defined the ratio $\alpha\equiv m_D/M\approx 0.93$. Equation~\eqref{Eq:PionExPot} corresponds to the result obtained by Baru \etal\ \cite{Baru:2011rs}.
An important feature of the given process is that exchanged pions can go on shell. This is the case whenever the denominator in Eq.~\eqref{Eq:PionExPot} vanishes. As a result, pion exchanges modify the \X\ width at NLO (see below).

\subsubsection{Partial Wave Projection}

Up to now, the amplitude involves arbitrary parities and total angular momenta \mbox{$\bs{J}\equiv \bs{L}+\bs{S}$} with total spatial angular momentum $\bs{L}$ and total spin $\bs{S}$. The \X, however, is a $J^P=1^+$ state. Thus, the total spin $S=1$ implies $L\in\{0,\,2\}$. We perform a respective partial wave projection of $T^{ij}$ by absorbing momentum dependences into scalar components $T_{LL';\,J}$ and angular dependences into projection operators of the form $P_{LL';\,J}^{ij}$ (see Appendix~\ref{Sec:AppendixPartWaveComps} for details). The functions $T_{LL';\,J}$ represent amplitudes connecting $D\bar{D}^\ast$ states with quantum numbers $L$ and $L'$, respectively, at total $J$. We can (schematically) write $T^{ij}=T_{0 0;\,1} P_{00;\,1}^{ij}+\cdots$ with $s$-wave projector $P_{00;\,1}^{ij}=\delta^{ij}$. The pion exchange potential $V^{ij}$ is expanded in the same fashion.

We aim at the calculation of $T_{0 0;\,1}$, which, through pion exchanges, is coupled to $T_{2 0;\,1}$. After projection, we may drop the subscript $J=1$ for convenience and find the scalar amplitude system
\begin{align}
	\label{Eq:ScalarAmps}
	\nonumber
	\left(\begin{matrix}
		T_{0 0}
		\vspace{1mm}\\
		T_{2 0}
	\end{matrix}\right)
	\left(p,\,p';\,E\right)\
	=&\ 
	-\left(\begin{matrix}
		C_0(\Lambda)+V_{0 0}+\mathcal{I}_\text{c}\vspace{1mm}
		\\
		V_{2 0}
	\end{matrix}\right)
	\left(p,\,p';\,E\right)
	\\
	&\hspace{-1cm}+4\pi\int_0^\Lambda
	\frac{\text{d}q\ q^2}{(2\pi)^3}\,
	\frac{
		\left(\begin{matrix}
			C_0(\Lambda)+V_{0 0}+\mathcal{I}_\text{c} & V_{0 2}
			\vspace{1mm}\\
			V_{2 0} & V_{2 2}
		\end{matrix}\right)
		\left(p,\,q;\,E\right)
	}{
		{G_\ast^\text{(f)}}^{-1}\Big(E-q^2/(2\mu_\ast)\Big)
	}
	\left(\begin{matrix}
		T_{0 0}\vspace{1mm}
		\\
		T_{2 0}
	\end{matrix}\right)
	\left(q,\,p';\,E\right).
\end{align}
The scalar components $V_{LL'}$ of the pion exchange potential are given in Appendix~\ref{Sec:AppendixPartWaveComps}. Note that the loop integral is in general divergent. Still, for a fixed momentum cutoff $\Lambda$, the system can be solved numerically for $T_{00}$.

\subsubsection{Numerical Renormalization}

For arbitrary cutoffs $\Lambda$ we tune $C_0(\Lambda)$ such that $T_{00}$ exhibits a pole at the complex energy
\begin{equation}
	\label{Eq:EX}
	E_X \equiv (\delta-\dX)-i\,\frac{\GX}{2}
\end{equation}
just below the $D\bar{D}^\ast$ threshold. More precisely, we fix the real part of $E_X$, i.e., the binding energy $\dX$ of the \X. Thereby, we obtain a prediction for the width $\GX$ as a function of $\dX$. At each order, $\GX$ should be independent of the cutoff as $\Lambda\rightarrow \infty$.

Note, that for large momenta, i.e., $p,\,q\gg (2 m_\pi |E|)^{1/2}$, the component $V_{00}$ of the pion exchange potential in Eq.~\eqref{Eq:PWPot00} approaches the constant $V_{00}^{(\infty)}\equiv -2/3\cdot g^2 \mu^2/m_\pi$. As a consequence, the loop integral in Eq.~\eqref{Eq:ScalarAmps} diverges as $\Lambda\rightarrow \infty$. This divergence is cured by the contact interaction $C_0(\Lambda)$: the curve $C_0(\Lambda)$ will be shifted by the amount $-V_{00}^{(\infty)}>0$ when $s$-wave pion exchanges enter the calculation. That is equivalent to introducing a counterterm $-V_{00}^{(\infty)}$ for pion exchanges.

\subsubsection{Application: $D^0\bar{D}^0\pi^0$ Production Rate}

\begin{figure}
	\begin{center}
		\includegraphics[scale=1]{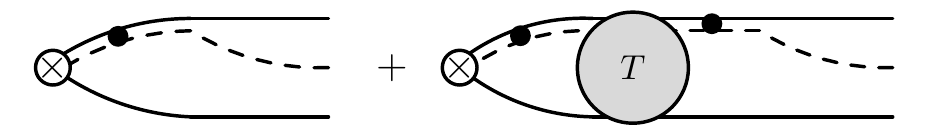}
	\end{center}
	\vspace{-5mm}
	\caption{\label{Fig:DDbarPiProd}Matrix element $i\,\mathcal{M}$ for $D\bar{D}\pi$ production. Short-range details are absorbed into a factor $\mathcal{F}$ indicated by a crossed circle. Each intermediate and final state is in a $C=+$ superposition.}
\end{figure}

The renormalized amplitude can be used to calculate the $D\bar{D}\pi$ production rate. This quantity can be measured and thus serves as an important link between theory and experiment. As done by Baru \etal\,, we consider a process in which the \X\ is produced at short ranges and subsequently decays to $D\bar{D}\pi$. As pointed out by Braaten and Lu \cite{Braaten:2007dw}, short-range details can be absorbed into a constant $\mathcal{F}$. The resulting differential rate is given by a phase space integral over the matrix element $\mathcal{M}$ depicted in Fig.~\ref{Fig:DDbarPiProd}. For outgoing particle momenta $\bs{p}_D,\,\bs{p}_{\bar{D}}$ and $\bs{p}_\pi=-(\bs{p}_D+\bs{p}_{\bar{D}})$, we obtain
\begin{align}
	\nonumber
	i\,\mathcal{M}^i(\bs{p}_D,\,\bs{p}_{\bar{D}};\, E)=&\ i
	\,g\,\frac{\mathcal{F}}{\sqrt{2}}\,
	\frac{(\alpha \bs{p}_D+\bs{p}_{\bar{D}})^i}{{G_\ast^\text{(f)}}^{-1}\left(E-p_D^2/(2\mu_\ast)\right)}
	\left(
		1-\int_0^\Lambda\frac{\text{d}q\,q^2}{(2\pi)^3}\,
		\frac{4\pi\,T_{00}(q,\,p_D;\,E)}{{G_\ast^\text{(f)}}^{-1}\left(E-q^2/(2\mu_\ast)\right)}
	\right)
	\\
	&\ +\left[\bs{p}_D\leftrightarrow \bs{p}_{\bar{D}}\right].
\end{align}
The $D\bar{D}\pi$ production rate for a system energy $E\in\mathbb{R}$ reads
\begin{equation}
	\label{Eq:ProdRate}
	\frac{\text{d}\Gamma}{\text{d}E}=
	\int\!\frac{\text{d}^3 p_D}{(2\pi)^3}
	\int\!\frac{\text{d}^3 p_{\bar{D}}}{(2\pi)^3}
	\ 
	2\pi\,\delta\!\left(E-\frac{p_D^2+p_{\bar{D}}^2}{2\mu}-\frac{\bs{p}_D\cdot\bs{p}_{\bar{D}}}{m_\pi}\right)
	\sum_i\left|\mathcal{M}^i(\bs{p}_D,\,\bs{p}_{\bar{D}};\, E)\right|^2.
\end{equation}

For consistency, the $D^\ast$ propagators in $\mathcal{M}$ will be chosen like the one entering the calculation of $T$. Moreover, since $\mathcal{F}$ is unknown, we have to normalize the rate. We follow Braaten and Lu by choosing the peak maximum in the $\dX=0$ rate to be $1$ \cite{Braaten:2007dw}.
The normalized line shapes will be independent of the cutoff as $\Lambda\rightarrow\infty$.

The rate exhibits a peak near the $D\bar{D}^\ast$ threshold representing the \X. In order to account for possible deviations from the pole parameters $\dX$ and $\GX$, the position of the peak maximum ($E_\text{max}$) and the full width at half maximum (FWHM) will be denoted by
\begin{equation}
	\label{Eq:PeakParameters}
	E_\text{max}\equiv \delta - \tilde{\delta}_X\,,
	\qquad
	\text{FWHM} \equiv \tilde{\Gamma}_{\!X}\,.
\end{equation}
This distinction will be of importance once $\dX$ becomes comparable to $\GX$.

\subsection{Momentum Scales}

The tiny binding energy $\dX$ introduces a new small energy scale. Equivalently, in terms of relative $D\bar{D}^\ast$ momenta, we find \textit{two} low-momentum scales
\begin{equation}
	\label{Eq:MomSc3Body}
	P_\ast\equiv \sqrt{2\mu_\ast \delta}\approx \unit[117]{MeV}\,,
	\qquad
	P_X\equiv \sqrt{2\mu_\ast |\dX|}\equiv \rho\,P_\ast\in \unit[{[0,\,20]}]{MeV}\,.
\end{equation}
The interval given in Eq.~\eqref{Eq:MomSc3Body} stems from the uncertainty range of $\dX$ in Eq.~\eqref{Eq:dX}. Note that the XEFT power counting does not distinguish between powers of $P_X$ and $P_\ast$ \cite{Fleming:2007rp}. Our scheme improves upon this point by counting them separately.

For convenience, we express the small ratio $\rho\equiv P_X/P_\ast$ in terms of $\kappa\sim 0.3$ by choosing $m\in\mathbb{Z}$ such that $\rho\in \kappa^m [\kappa^{1/2},\,\kappa^{-1/2})$. For $m=2$, this interval corresponds almost exactly to the positive part of the uncertainty range of $\dX$, i.e.,
\begin{equation}
	\dX=\delta\,\rho^2\in \delta\, \kappa^4\left[\kappa,\,\kappa^{-1}\right)
	\approx \left[0.017,\,0.190\right).
\end{equation}
Therefore, we count $\rho\sim \kappa^2\sim 0.09$ ($P_X\sim \unit[11]{MeV}$) in the following. Note that the central value $\dX\sim \delta\, \kappa^4 \approx \unit[0.057]{MeV}$ is exactly one-third of the upper interval limit $0.190$. Thus, we systematically favor small values of $\dX$. This choice is in line with the fact that the experimental centroid of $\dX$ lies close to zero.

The high-momentum scale of the three-body system is expected to lie in the chiral breakdown regime $\Lambda_\chi\sim \unit[500]{MeV}$ of Heavy Hadron Chiral Perturbation Theory. In this region, also pion production takes place, i.e., $(2\mu_\ast m_\pi)^{1/2}\approx \unit[510]{MeV}$. Note, that in XEFT the hard scale is taken to be the pion mass itself. However, even for relative momenta of order $m_\pi\gtrsim P_\ast$ charm mesons are nonrelativistic, and also relativistic pion corrections are small (see Appendix~\ref{Sec:AppendixRelativistics}).

\subsection{$\bs{X(3872)}$ Width at LO}

\begin{figure}
	\begin{center}
		\includegraphics[scale=1]{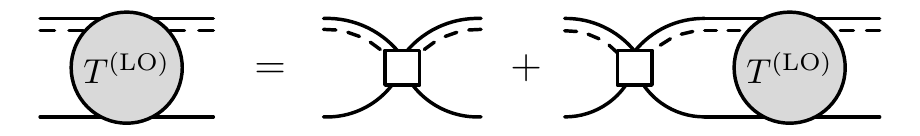}
	\end{center}
	\vspace{-0.5cm}
	\caption{\label{Fig:ScalarAmpLO}Amplitude for calculating the width $\GX$ at LO.}
\end{figure}

We will see below that width contributions from pion exchanges, $D^\ast$ propagator corrections and charged mesons are subleading. Thus, the LO width can be obtained by iterating the leading-order $D^\ast$ propagator of Eq.~\eqref{Eq:GStarLO} alongside $C_0$. The corresponding amplitude $T_{00}^\text{(LO)}$ is shown in Fig.~\ref{Fig:ScalarAmpLO} and yields
\begin{equation}
	T_{00}^\text{(LO)}(E)=\left[
		-{C_0^\text{(LO)}}^{-1}(\Lambda)
		-\frac{\mu_\ast}{2\pi}\left(
			\frac{2}{\pi}\Lambda-\sqrt{2\mu_\ast(E_\ast-E-i\epsilon)}
			+\mathcal{O}(\Lambda^{-1})
		\right)
	\right]^{-1}.
\end{equation}
We demand $0\equiv {T_{00}^\text{(LO)}}^{-1}(E_X^{\text{(LO)}})$ with $E_X^{\text{(LO)}}\equiv (\delta-\dX)-i\,\GX^{\text{(LO)}}/2$ and $C_0^{\text{(LO)}}(\Lambda)\in\mathbb{R}$, yielding
\begin{equation}
	\label{Eq:LOResultsGXC0}
	\GX^{\text{(LO)}}=\Gast\,,
	\qquad
	{C_0^{\text{(LO)}}}^{-1}(\Lambda)
	= -\frac{\mu_\ast}{2\pi}
	\left(
		\frac{2}{\pi}\Lambda
		-\sqrt{2\mu_\ast\dX}
		+\mathcal{O}(\Lambda^{-1})
	\right).
\end{equation}
As expected, the LO width is given by the full $D^\ast$ width, independently of $\dX$.

The renormalized amplitude reads
\begin{equation}
	\label{Eq:T00LO}
	T_{00}^\text{(LO)}(E)=-\frac{2\pi}{\mu_\ast}\,\left[
		\sqrt{2\mu_\ast\delta_X}-\sqrt{2\mu_\ast(E_\ast-E-i\epsilon)}
	\right]^{-1}
	\equiv \text{reg}+\frac{Z^{\text{(LO)}}}{E-E_X^{\text{(LO)}}+i\epsilon}\,,
\end{equation}
where ``reg'' stands for terms regular at the LO pole position $E_X^{\text{(LO)}}=E_\ast-\dX$. Note, that our LO amplitude almost recovers the zero-range result by Braaten and Lu \cite{Braaten:2007dw}. They used an energy-dependent $D^\ast$ width instead of the constant one in Eq.~\eqref{Eq:T00LO}. This energy dependence can be neglected at LO. The value of the LO residue,
\begin{equation}
	\label{Eq:LOResidue}
	Z^\text{(LO)}=-\frac{2\pi}{\mu_\ast^2}\,\sqrt{2\mu_\ast\dX}\,,
\end{equation}
is of great importance for the \X\ width: All sub-leading width contributions will at least be proportional to $Z^\text{(LO)}$ and therefore to the small momentum $P_X=(2\mu_\ast|\dX|)^{1/2}\sim \unit[11]{MeV}$.

\subsection{NLO Corrections to the Width}

In the following, we verify the LO nature of $\GX^\text{(LO)}=\Gast$. Similar to the two-body sector, the expansion of the width will be in $\kappa^2\sim 0.09$. Self-energy corrections, charged mesons, and pion exchanges between $s$-wave states will enter at NLO ($\kappa^2$). Note that the predictive power of our EFT is limited by the experimental uncertainty levels. The largest such uncertainties come from $\Gc=\Gamma[D^{+\,\ast}\rightarrow D^0\pi^+ + D^+\pi^0]$ and $\mathcal{B}=\GDg/(\GDPi+\GDg)$ and are of the order $3\%\approx\kappa^3$. Thus, we expect our NLO results to be reliable. We remark that, in principle, there are also NLO corrections to the real part of the complex energy $E_X$.  In our renormalization scheme, however, the real part of $E_X$ is kept fixed by properly readjusting $C_0$ as explained below.

\subsubsection{Power Counting}

Let $q^\mu=(q^0,\,\bs{q})$ be a relative $D\bar{D}^\ast$ four-momentum. Loop integrations are counted non-relativistically, i.e., $\text{d}^4 q \sim q^5$ with $q=|\bs{q}|$. We investigate the $D\bar{D}^\ast$ amplitude in the vicinity of the \X\ pole, i.e., at $E\approx E_\ast$. In this region, the $D$ propagator $i\,G_D$ as well as the LO $D^\ast$ propagator $i\,G_\ast$ count like $q^{-2}$. The propagator of an exchanged pion, however, depends both on the incoming and outgoing relative momentum $p_\text{in}$ and $p_\text{out}$, as can be seen from Eq.~\eqref{Eq:PionExPot}. Furthermore, it is suppressed by the small mass ratio
\begin{equation}
	r\equiv \mu/\mu_\ast\approx 0.13\,.
\end{equation}
Consequently, we count $i\,G_\pi\sim r\left( \max\{p_\text{in}^2,\,p_\text{out}^2\}\right)^{-1}$.

Finally, in this section the coupling $g$ has to be expressed in terms of the reduced mass $\mu_\ast$ and the momentum scales $P_\ast,\,P_X$, yielding $1/(2\mu_\ast g^2)\approx 0.8\,P_\ast\,\mu_\ast/(2\pi)$. Therefore, we count $g\sim P_\ast^{-1/2}$ in Feynman diagrams.

\subsubsection{Width Estimation Strategy}

We consider an arbitrary $D\bar{D}^\ast$ interaction $\mathcal{I}$ other then $C_0$. If resummed to all orders, it shifts the LO pole position and residue to $E_X^{(\mathcal{I})}\equiv E_X^\text{(LO)}+\Delta E_X^{(\mathcal{I})}$ and $Z^{(\mathcal{I})}\equiv Z^\text{(LO)}+\Delta Z^{(\mathcal{I})}$, respectively. The new amplitude $T_{00}^{(\mathcal{I})}\equiv T_{00}^\text{(LO)}+\Delta T_{00}^{(\mathcal{I})}$ can be expanded at LO pole as follows:
\begin{align}
	\label{Eq:AmpShift}
	T_{00}^{(\mathcal{I})}
	=&\ \text{reg}+\frac{Z^{(\mathcal{I})}}{E-E_X^{(\mathcal{I})}+i\epsilon}
	=\text{reg}+\frac{Z^{(\mathcal{I})}}{E-E_X^\text{(LO)}+i\epsilon}
	+\frac{Z^{(\mathcal{I})}\Delta E_X^{(\mathcal{I})}}{\left(E-E_X^\text{(LO)}+i\epsilon\right)^2}
	+\cdots\,.
\end{align}
By comparison with the generic form
\begin{equation}
	\Delta T_{00}^{(\mathcal{I})}\equiv a^{(\mathcal{I})}\,
	T_{00}^\text{(LO)}
	+ b^{(\mathcal{I})} \left(T_{00}^\text{(LO)}\right)^2+\cdots
\end{equation}
we identify the shifts $\Delta E_X^{(\mathcal{I})}=Z^\text{(LO)}b^{(\mathcal{I})}/(1+a^{(\mathcal{I})})$ and $\Delta Z^{(\mathcal{I})}=Z^\text{(LO)}\,a^{(\mathcal{I})}$. The coefficients $a^{(\mathcal{I})}$ and $b^{(\mathcal{I})}$ can be determined diagrammatically.
Jansen \etal\ have used this procedure to calculate $\dX$ at NLO in XEFT \cite{Jansen:2013cba}. Our renormalization scheme, in contrast, keeps $\dX$ fixed by readjusting $C_0$. In other words, we resum an appropriate correction term $\Delta C_0^{(\mathcal{I})}$ in addition to $\mathcal{I}$ that cancels the real part of $\Delta E_X^{(\mathcal{I})}$.
	
Note that $a^{(\mathcal{I})}$ and $b^{(\mathcal{I})}$ can be momentum dependent, while the expression $\Delta E_X^{(\mathcal{I})}\propto b^{(\mathcal{I})}/(1+a^{(\mathcal{I})})$ must be a number. In Ref.~\cite{Jansen:Diss} it was shown that this momentum dependence indeed cancels at NLO in XEFT. More generally, it follows from the momentum independence of $T_{00}^\text{(LO)}$, that $b^{(\mathcal{I})}/(1+a^{(\mathcal{I})})=\bar{b}^{(\mathcal{I})}$, where $\bar{b}^{(\mathcal{I})} (T_{00}^\text{(LO)})^2$ contains all diagrams with interactions $\mathcal{I}$ between the two LO amplitudes, like in Fig.~\ref{Fig:WidthContributions}. Such diagrams are always momentum independent. Thus, we may write the width shift in the form
\begin{equation}
	\label{Eq:WidthEstimation}
	\Delta \GX^{(\mathcal{I})}/2=
	-Z^\text{(LO)}\,\operatorname{Im}\bar{b}^{(\mathcal{I})}\,.
\end{equation}
We see from Eq.~\eqref{Eq:WidthEstimation} that each correction to $\GX^\text{(LO)}$ is proportional to $Z^\text{(LO)}$ and thus to the small binding momentum $P_X = \rho\, P_\ast$ with $\rho\sim \kappa^2$. We can now verify the power counting order of $D^\ast$ propagator corrections, pion exchanges, and charged mesons in the following way:
\begin{enumerate}
	\item Identify all diagrams induced by $\mathcal{I}$ that contribute to $\bar{b}^{(\mathcal{I})}\,(T_{00}^\text{(LO)})^2$.
	\item Determine their overall scaling by investigating loop momenta at both $P_X$ and $P_\ast$. Imaginary parts from on-shell pion exchanges are to be investigated separately.
	\item Estimate $\Delta\GX^{(\mathcal{I})}$ using Eq.~\eqref{Eq:WidthEstimation}.
\end{enumerate}

\subsubsection{Propagator Corrections}

\begin{figure}
\begin{center}
	\subfigure[]{
		\includegraphics[scale=1]{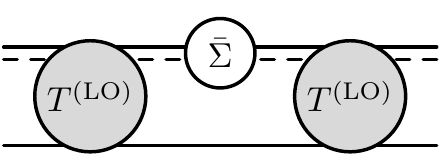}
	}
	\subfigure[]{
		\includegraphics[scale=1]{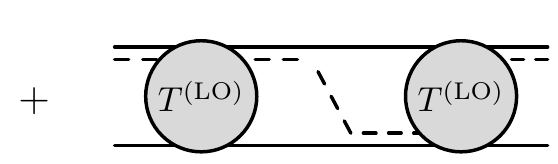}
	}
	\end{center}
	\vspace{-0.5cm}
	\caption{\label{Fig:WidthContributions} Pion interaction diagrams contributing to $\bar{b}^{(\mathcal{I})}\,(T_{00}^\text{(LO)})^2$ at NLO. The width shift due to single self-energy corrections is proportional to diagram (a), while diagram (b) determines the shift due to one-pion exchanges.}
\end{figure}

The width shift $\Delta \GX^{(1\Sigma)}$ due to single self-energy corrections in the $D^\ast$ propagator is proportional to the one-loop diagram in Fig.~\ref{Fig:WidthContributions}(a), evaluated at $E=E_X^{\text{(LO)}}$. Let $q^\mu$ be the loop's four-momentum. The $D^\ast$ center-of-mass energies for $q\sim P_X$ and $q\sim P_\ast$ lie in the regions $\Ecm\approx E_\ast$ and $\Ecm\approx 0$, respectively. As discussed in Sec.~\ref{Sec:DPi}, self-energy corrections $-i\,\bar{\Sigma}\cdot i\,G_\ast$ are of the order $\chi^2+i\,\chi$ with $\chi=\GDPi/(2\delta)$ in both regions. The two remaining propagators and the integral measure contribute a factor $q$. Thus, the main contribution to the integral stems from the region $q\sim P_\ast$, and we find $\bar{b}^{(1\Sigma)}\sim \left(\chi^2+i\,\chi\right)P_\ast$. That yields
\begin{equation}
	\Delta\GX^{(1\Sigma)}/2 =
	-\,Z^\text{(LO)}\operatorname{Im}\bar{b}^{(1\Sigma)}
	\sim P_X\,\chi\,P_\ast
	\sim \rho\,\chi\,\delta
	\sim \rho\,\GX^\text{(LO)}/2\,,
\end{equation}
with $Z^\text{(LO)}\propto P_X=\rho\,P_\ast$ [see Eq.~\eqref{Eq:LOResidue}] and $\delta\sim P_\ast^2$ [see Eq.~\eqref{Eq:MomSc3Body}]. From this estimation we expect that $\Delta\GX^{(1\Sigma)}$ corrects the \X\ width at NLO ($\rho\sim\kappa^2$).

Apart from the scaling, $\operatorname{Im}\bar{b}^{(1\Sigma)}$ determines also the sign of $\Delta\GX^{(1\Sigma)}$. After performing the $q_0$ integral in Fig.~\ref{Fig:WidthContributions}(a) and counting all $\pm i$ factors from Feynman rules, we can symbolically write $\operatorname{sign}\Delta\GX^{(1\Sigma)}=\operatorname{sign}\operatorname{Im}\bar{b}^{(1\Sigma)}=-\operatorname{sign}\operatorname{Im}[G_\ast\cdot\bar{\Sigma}\cdot G_\ast]$. At $E=E_X^{\text{(LO)}}$, we have $G_\ast<0$ and for $q\sim P_\ast$, the correction $\bar{\Sigma}\cdot G_\ast\approx \Sigma(E_\ast)/E_\ast$ has a negative imaginary part. Thus, the width shift $\Delta\GX^{(1\Sigma)}$ is negative and decreases the overall width.

Compared to the self-energy correction, the imaginary parts of all other propagator corrections are suppressed by orders of $\kappa^2$; see Eq.~\eqref{Eq:PropCorrsAtRes1} and \eqref{Eq:GFullComplexFactorized}. Thus, they do not enter before N${}^2$LO ($\kappa^4$). Relativistic corrections to the $D^\ast$ propagator are even more suppressed (see Appendix~\ref{Sec:AppendixRelativistics} for details).

\subsubsection{Pion Exchanges}

Next, we consider the resummation of one-pion exchanges.
The factor $\bar{b}^{(1\pi)}$ is given by the two-loop diagram in Fig.~\ref{Fig:WidthContributions}(b) with loop four-momenta $q^\mu,\,s^\mu$. Its absolute value can be estimated like above. Recalling that the pion propagator scales like $G_\pi\sim r \left(\max\{q^2,\,s^2\}\right)^{-1}$ with $r=\mu/\mu_\ast$ and the vertices count like $g\max\{q,\,s\}$ with $g\sim P_\ast^{-1/2}$, we obtain the overall product $r\, q\,s/P_\ast$. Thus, the absolute value of the integral is governed by loop momenta $q\sim s\sim P_\ast$ yielding $|\bar{b}^{(1\pi)}|\sim r\,P_\ast$.

The integral's imaginary part, however, scales differently for it appears only if the pion goes on shell. This restriction imposes a condition on the angle $\cos\theta\equiv\bs{e}_q\cdot\bs{e}_s$. Due to Eq.~\eqref{Eq:PionExPot}, it has to behave like
\begin{equation}
	\label{Eq:AngleCondition}
	\cos\theta=\frac{m_\pi}{qs}\left(E-\frac{q^2+s^2}{2\mu}\right)
	\sim \frac{r\,P_\ast^2 - q^2 - s^2}{2\,qs}\,,
\end{equation}
with $m_\pi\sim \mu$. This relation has no solution for $P_X\sim q\ll s\sim P_\ast$ or vice versa, because the right-hand side falls outside the interval $[-1,\,1]$. Thus, on-shell pions require $q\sim s$. In this case, Eq.~\eqref{Eq:AngleCondition} yields $\cos\theta\lesssim +1$ for small momenta $q\sim s\gtrsim r^{1/2}P_\ast/2\approx 2\,P_X$. The corresponding overall factor is of size $q\,r\,P_\ast^{-1} s\sim r^2 P_\ast/4$. Naively, one would expect that the contribution at $q\sim s\sim P_\ast$ scales like $P_\ast\, r\, P_\ast^{-1} P_\ast= r\,P_\ast$, which is much larger. However, in this region we have $\cos\theta\approx -1$, which leads to a near-cancellation of the product of the two pion vertices: for $\bs{q}=-\bs{s}$, the pion exchange potential becomes proportional to the suppression factor $(1-\alpha)^2\approx r^2/4$; see Eq.~\eqref{Eq:PionExPot}. Therefore, the imaginary part in this region is subleading, and we find $\bar{b}^{(1\pi)}\sim \left(r+i\,r^2/4\right)P_\ast$. From Eq.~\eqref{Eq:WidthEstimation} we obtain the estimation
\begin{equation}
	\Delta\GX^{(1\pi)}/2= -Z^\text{(LO)}\operatorname{Im}\bar{b}^{(1\pi)}
	\sim \rho\,\underbrace{
		\left(r^2/4\right)
	}_{\approx 1.7\,\chi}\,\delta
	\sim \rho\,\GX^\text{(LO)}/2\,.
\end{equation}
Thus, the correction enters at NLO ($\rho\sim\kappa^2$) as well.

As done for the self-energy contribution, we can infer the sign of $\Delta\GX^{(1\pi)}$ from the diagram in Fig.~\ref{Fig:WidthContributions}(b). We note that integrating over the pion propagator produces a negative imaginary part, which follows from the $+i\epsilon$ prescription.  Moreover, the product of the two pion vertices is always negative. Taking into account all remaining phase factors, we obtain $\operatorname{sign}\Delta\GX^{(1\pi)}=\operatorname{sign}\operatorname{Im}\bar{b}^{(1\pi)}=+1$. Thus, single pion exchanges increase the width. In fact, we will see that this leads to a near-cancellation of the self-energy corrections.

Relativistic corrections to the pion propagator enter at N${}^2$LO ($\kappa^4$). Further details are given in Appendix~\ref{Sec:AppendixRelativistics}. Moreover, contributions from multi-pion exchanges are at least suppressed by additional factors of $r\gtrsim \kappa^2$. From this observation, we draw two conclusions. First, we may resum all pion exchanges between $s$-waves at NLO. Second, contributions from $d$-wave $D\bar{D}^\ast$ states are of the order N${}^2$LO as they involve at least two pion exchanges.

\subsubsection{Charged Mesons}

At NLO, charged states $(D^+D^{-\ast}+D^-D^{+\ast})/\sqrt{2}$ cannot be neglected. In Ref.~\cite{Baru:2011rs} they have been included to all orders via charged pion exchanges and $s$-wave contact interactions. However, charged pion exchanges (just as neutral ones) involve additional suppression factors of order $r=\mu/\mu_\ast\ll 1$, which makes them subleading.
\begin{figure}
	\begin{center}
		\includegraphics[scale=1]{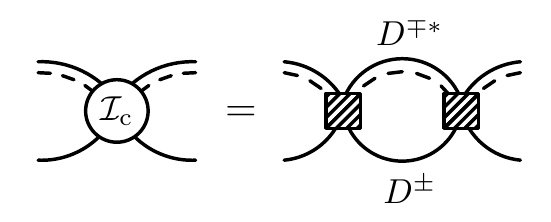}
	\end{center}
	\vspace{-1cm}
	\caption{\label{Fig:Charged}Effective $D\bar{D}^\ast$ interaction $-i\,\mathcal{I}_\text{c}$ for the leading contribution of intermediate charged states $(D^+D^{-\ast}+D^-D^{+\ast})/\sqrt{2}$. The shaded vertex is defined via Eq.~\eqref{Eq:ChargedInteraction}.}
\end{figure}
Instead of a nonperturbative treatment, we include charged $D^{(\ast)}$ mesons through the effective interaction $-i\,\mathcal{I}_\text{c}(E)$ appearing in Eq.~\eqref{Eq:3BodyAmp}. It contains only contact interactions. Due to isospin symmetry, the vertex connecting a neutral and a charged $C=+$ combination exhibits a factor $2$ compared to the vertex between neutral pairs \cite{Baru:2011rs}. However, it may not contain a counterterm for pion exchanges since we exclude charged pions. Therefore, whenever neutral pion exchanges enter the computation, we subtract the counterterm $-V_{00}^{(\infty)}=2/3\cdot g^2 \mu^2/m_\pi$ from the vertex. The resulting interaction is shown diagrammatically in Fig.~\ref{Fig:Charged}. It reads
\begin{equation}
	\label{Eq:ChargedInteraction}
	-i\,\mathcal{I}_\text{c}(E)
	\equiv i\,\left(2\big(C_0+V_{00}^{(\infty)}\big)\right)^2\,
	\frac{\mu_{\ast,\text{c}}}{2\pi}\left(
		\frac{2}{\pi}\Lambda
		-\sqrt{
			2\mu_{\ast,\text{c}}
			\left(\delta +\nu -E-i\epsilon\right)
		}+\mathcal{O}(\Lambda^{-1})
		\right),
\end{equation}
with $\mu_{\ast,\text{c}}\equiv(m_{D^+}^{-1}+m_{D^{+\ast}}^{-1})^{-1}\approx\mu_\ast$
and $\nu$ as defined in Eq.~\eqref{Eq:DeltaM}.
For more details on the charged meson propagators in Fig.~\ref{Fig:Charged}, see Appendix~\ref{Sec:AppendixCharged}.

The perturbative inclusion of charged mesons has several advantages. First of all, we do not need to introduce an additional scattering channel, keeping the system matrix small. Furthermore, the system becomes renormalizable for arbitrary $\Lambda$. Finally, the effect of the interaction on $\GX$ is analytically solvable if pion exchanges and propagator corrections are switched off. We iterate $-i\,\mathcal{I}_\text{c}(E)$ alongside $-i\,C_0$ (with $V_{00}^{(\infty)}\equiv 0$) and set the pole to $E_X^\text{(LO)}-i\,\Delta\GX^{(\mathcal{I}_\text{c})}/2$. Again, we demand $C_0(\Lambda)\in\mathbb{R}$ and choose the one solution of $C_0$ that recovers the LO expression in the limit $\mathcal{I}_\text{c}\rightarrow 0$. This procedure yields
\begin{equation}
	\label{Eq:ChargedCorrection}
	\Delta\GX^{(\mathcal{I}_\text{c})}
	=\underbrace{
		-\sqrt{\frac{\mu_{\ast,\text{c}}}{\mu_\ast}}\,
		\left(
			1+\frac{\mu_\ast}{\mu_{\ast,\text{c}}}\,\frac{
				1+\sqrt{
					1
					+16\,\frac{\mu_{\ast,\text{c}}}{\mu_\ast}
				}
			}{8}
		\right)^{-1}
	}_{\approx\ -0.6}\ 
	\sqrt{\frac{\dX}{\nu}}\ \Gast\ 
	\left(
		1+\mathcal{O}\left(
			\sqrt{\frac{\dX}{\nu}}
		\right)
	\right).
\end{equation}
We see that charged mesons lower the \X\ width, which is in line with the findings by Baru \etal~\cite{Baru:2011rs}.

Compared to the binding energy $\dX$, the energy difference $\nu\approx \unit[8]{MeV}$ of the $D^{\pm\ast}D^\mp$ threshold to the \X\ is large. It corresponds to a $D\bar{D}^\ast$ momentum of the order $(2\mu_\ast\nu)^{1/2}\sim (2\mu_\ast\delta)^{1/2}=P_\ast$ due to $\nu\sim \delta$. Therefore, the width correction induced by the charged meson loop is of order NLO ($\rho\sim \kappa^2$). Similarly, contributions of multiple charged meson loops are suppressed by $\rho^2\sim \kappa^4,\,\rho^3\sim\kappa^6,\,$ etc., and do not enter before N${}^2$LO. This observation verifies the perturbative nature of charged mesons in the \X.

\subsection{Summary: Inputs and Outputs of the EFT}

We conclude this section by summarizing all EFT inputs and predictions in the two- and three-body sector up to NLO. They are listed in the Table~\ref{Tab:InputsOutputs}.

\begin{table}[h!]
	\caption{\label{Tab:InputsOutputs}Inputs and outputs of the EFT up to NLO.}
	\begin{tabular}{lcccc}
		\hline\hline
		& \multicolumn{2}{c}{\bf Two-Body system} & \multicolumn{2}{c}{\bf Three-Body system}
		\\
		& \bf \ \ Inputs\ \  & \bf \ \ Outputs & \bf \ \ Inputs\ \  & \bf \ \ Outputs\ \ 
		\\\hline
		\rule{0pt}{15pt} \bf LO ($\kappa^0$)\ \ & \ \ $\delta,\,\delta_{+0},\,\delta_{++},\,\Gc$\ \ 
		& \ \ $g^2,\,\GDPi,\,a_1^{-1},\,r_1/2$\ \ 
		& \ \ $\dX$\ \ 
		& \ \ $\GX,\,\text{d}\Gamma/\text{d} E\ (\text{with }\tilde{\delta}_X,\,\tilde{\Gamma}_{\!X})$\ \ 
		\\
		& $\mathcal{B}$ & $\GDg$ & &
		\\\hline
		\rule{0pt}{15pt} \bf NLO ($\kappa^2$)\ \ & (see LO) & (see LO) & \ \ $\nu$\ \  & (see LO)
		\\
		\hline\hline
	\end{tabular}
\end{table}

In the two-body system, we have used the mass splittings $\delta,\,\delta_{+0},\,\delta_{++}$ and the pionic decay width $\Gc$ of the $D^{+\ast}$ to determine the coupling $g$ (see Appendix~\ref{Sec:AppendixCoupling}) and further the width $\GDPi$ and the threshold parameters $a_1^{-1}$ and $r_1/2$. Subsequently, the radiative decay width $\GDg$ has been obtained from $\GDPi$ by taking the branching ratio $\mathcal{B}$ as additional input. All parameters are renormalized in the MS scheme. Note that the two-body predictions do not change from LO to NLO. The reason is that the LO $D^\ast$ propagator already contains the full $D^\ast$ width and the NLO self-energy correction involves no new parameters.

The three-body system can be renormalized using the coupling $C_0(\Lambda)$. The binding energy $\dX$ serves as renormalization condition at both LO and NLO, and at NLO, also the mass splitting $\nu$ between the neutral and charge thresholds is needed. Thereby, we obtain $\GX$ and the production rate $\text{d}\Gamma/\text{d} E$ as functions of $\dX$. Let us stress again that the physical value of $\dX$ is not precisly known. We will, however, see that there are one-to-one relations between $\dX$ and both the production rate's peak width $\tilde{\Gamma}_{\!X}$ and maximum position $\delta-\tilde{\delta}_{X}$ [see Eq.~\eqref{Eq:PeakParameters}]. They can be inverted in order to predict $\dX$ from the experimentally measured line shape.

\section{Results}
\label{Sec:Results}

In this section, we present numerical results for $\GX$ up to NLO ($\kappa^2$). As argued above, N${}^2$LO contributions ($\kappa^4$) would involve higher-order propagator corrections, intermediate $d$-waves, iterations of the charged meson interaction and relativistic corrections. For illustration, we explicitly calculate the effect of $d$-wave states and show that is of order N${}^2$LO. Moreover, we show that the system is renormalizable for arbitrary cutoffs. Afterward, we calculate the line shape of the \X\ in $D\bar{D}\pi$ production. We show that the peak's maximum position and line width can only be identified with the pole position if $\dX>\GX$ and if the detector resolution is sufficiently high.

\subsection{{$\bs{X(3872)}$} Width}

\begin{figure}
	\begin{center}
		\includegraphics[scale=1]{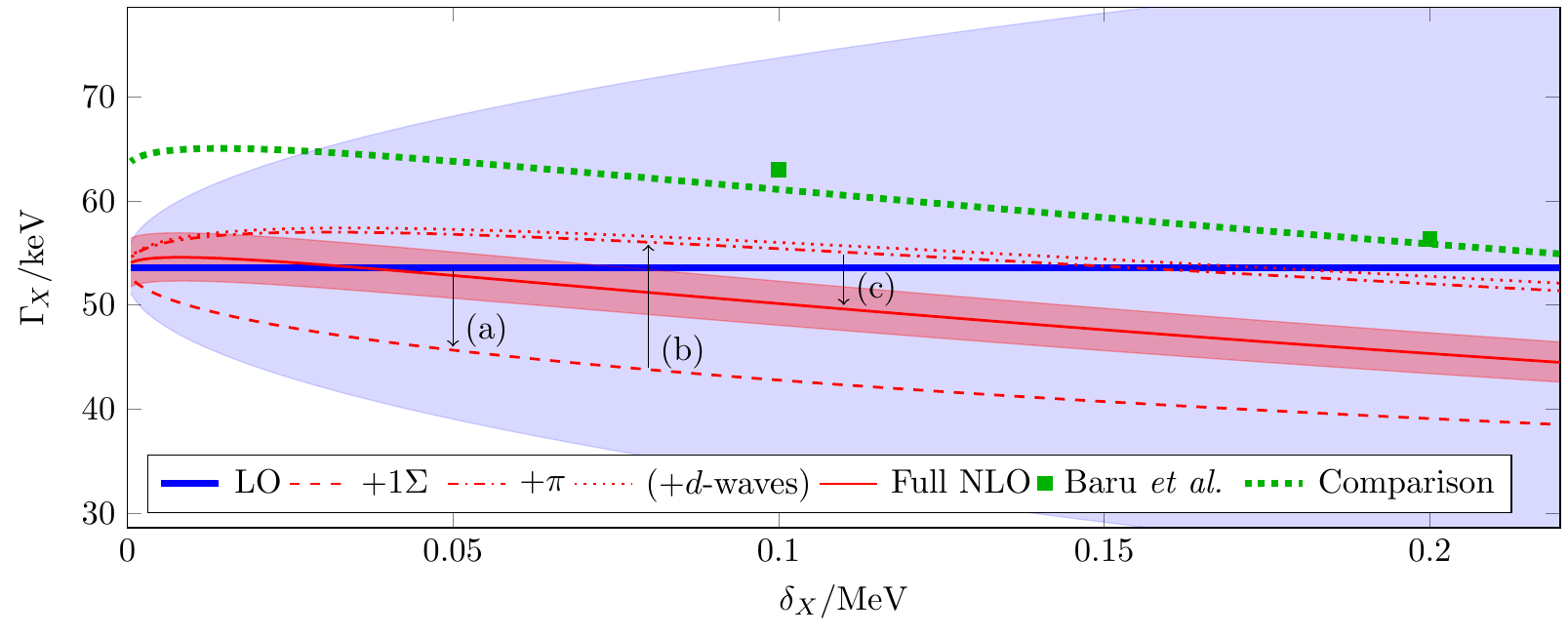}
	\end{center}
	\vspace{-1cm}
	\caption{\label{Fig:GX}Width $\GX$ as a function of $\dX$ up to NLO compared to the results of Ref.~\cite{Baru:2011rs}. The three arrows indicate the NLO corrections (a) $\Delta\GX^{(1\Sigma)}$, (b) $\Delta\GX^{(\pi)}$ and (c) $\Delta\GX^{(\mathcal{I}_\text{c})}$.}
\end{figure}

In order to assess our power counting predictions and to demonstrate the convergence of the scheme, we compare calculations at LO and NLO.
At LO, we solve the system depicted in Fig.~\ref{Fig:ScalarAmpLO}, which must yield $\GX^\text{(LO)}=\Gast$. All subleading corrections are expected to be at least proportional to $\rho=(\dX/\delta)^{1/2}$; see above. We may thus obtain an LO uncertainty band by shifting the $D^\ast$ width by $\pm 3\,\rho\,\Gast$. Thereby, we allow for a possible numerical coefficient. On top, we take into account the experimental uncertainties of $\Gc$ and $\mathcal{B}$ by varying \mbox{$g^2\in\unit[{[3.40,3.56]\cdot 10^{-8}}]{MeV^{-3}}$} and \mbox{$\mathcal{B}\in\unit[{[34.4,36.2]}]{\%}$}. The numerical results are presented in Fig.~\ref{Fig:GX}. The LO width, shown as a (blue) bold line, is indeed independent of $\dX$ and given by $\Gast=\unit[53.6]{keV}$. At $\dX=\unit[57]{keV}$ (i.e., $\rho=\kappa^2$) the LO band yields an uncertainty of about $\pm \unit[15]{keV}$.

At NLO, we add the three contributions step by step. First, we insert single self-energy corrections in the $D^\ast$ propagator as shown in Fig.~\ref{Fig:PropCorrs}. The resulting shift, shown as a (red) dashed line in Fig.~\ref{Fig:GX} shows a $\rho\propto\dX^{1/2}$ dependence as expected and lies within the LO band. Next, we introduce pion exchanges between relative $s$-wave states. As expected, the corresponding width shift is of the same order as the previous one but has the opposite sign.
As a consequence, at \mbox{$\dX=\unit[57]{keV}$}, we obtain a small overall shift of $+\unit[3]{keV}$ compared to the LO width. The influence of intermediate $d$-waves is expected to be of order $\kappa^4\,\Gast\approx \unit[0.44]{keV}$ (N${}^2$LO). This estimation is perfectly confirmed by the numerical result (red dotted line), which lies only $\unit[0.5]{keV}$ above the previous one. We conclude that $d$-waves are negligible at NLO and exclude them from all following calculations. The full NLO width, shown as a (red) solid line, is obtained by taking into account the charged meson loop. Remarkably, the overall NLO correction at $\dX=\unit[57]{keV}$ lies only $\unit[1.1]{keV}$ below the LO result. Moreover, a variation of experimental inputs yields an NLO uncertainty band of size $\pm \unit[2.2]{keV}\sim\kappa^3\,\Gast$, which surrounds the LO curve for small $\dX$. Thus, the simple analytic LO result lies 
within the NLO band up to $\dX\approx\unit[75]{keV}$.
In summary, all results are in very good agreement with the power counting predictions. Our full NLO prediction for the width at $\dX=\unit[57]{keV}$ reads
\begin{equation}
	\label{Eq:NLOWidthPrediction}
	\GX^\text{(NLO)}=\unit[(52.5\pm 2.2)]{keV}\,.
\end{equation}

It is instructive to compare the NLO prediction to the coupled channel results of Baru \etal\ \cite{Baru:2011rs}; see the (green) squares in Fig.~\ref{Fig:GX}. Taking into account the $D^\ast$ width $\Gast^\text{(Baru)}=\unit[63]{keV}$ used in Ref.~\cite{Baru:2011rs}, we obtain the (green) bold-dotted curve in Fig.~\ref{Fig:GX}. Indeed, both approaches agree very well for the same input parameters. Since the $D^\ast$ self-energy was treated non-perturbatively in  Ref.~\cite{Baru:2011rs}, this agreement
provides strong evidence for the subleading natures of the $D^\ast$ self-energy, as well as intermediate $d$-wave states, charged meson states in general, and charged pion exchanges specifically. Thus, we conclude that our power counting scheme exhibits quick convergence. At second glance, one sees a minor deviation of about $\unit[2]{keV}$ at $\dX=\unit[100]{keV}$. It indicates the beginning influence of threshold effects which blurr the \X\ peak in $D\bar{D}\pi$ production for small $\dX$. While deviations of the peak from a Breit-Wigner shape are negligible for the $\dX$ investigated in Ref.~\cite{Baru:2011rs}, they have to be accounted for in the region $\dX<\unit[100]{keV}$.

Let us emphasize at this point that our EFT is based on the molecular picture of the \X\,, which decays to $D\bar{D}\pi$ or $D\bar{D}\gamma$. Contributions from other decay channels might have a significant impact on $\GX$\,. Their inclusion, however, goes beyond the scope of this work and has to be addressed in the future. Moreover, note that the uncertainty given in Eq.~\ref{Eq:NLOWidthPrediction} relies on certain scaling assumptions for higher-order $D\pi$ terms as discussed in Sec.~\ref{Sec:DPi}. These assumptions represent a scenario of minimal fine-tuning. Although unlikely, further fine-tunings could thus invalidate the developed power counting.

\subsection{Contact Interaction}

\begin{figure}
	\begin{center}
		\includegraphics[scale=1]{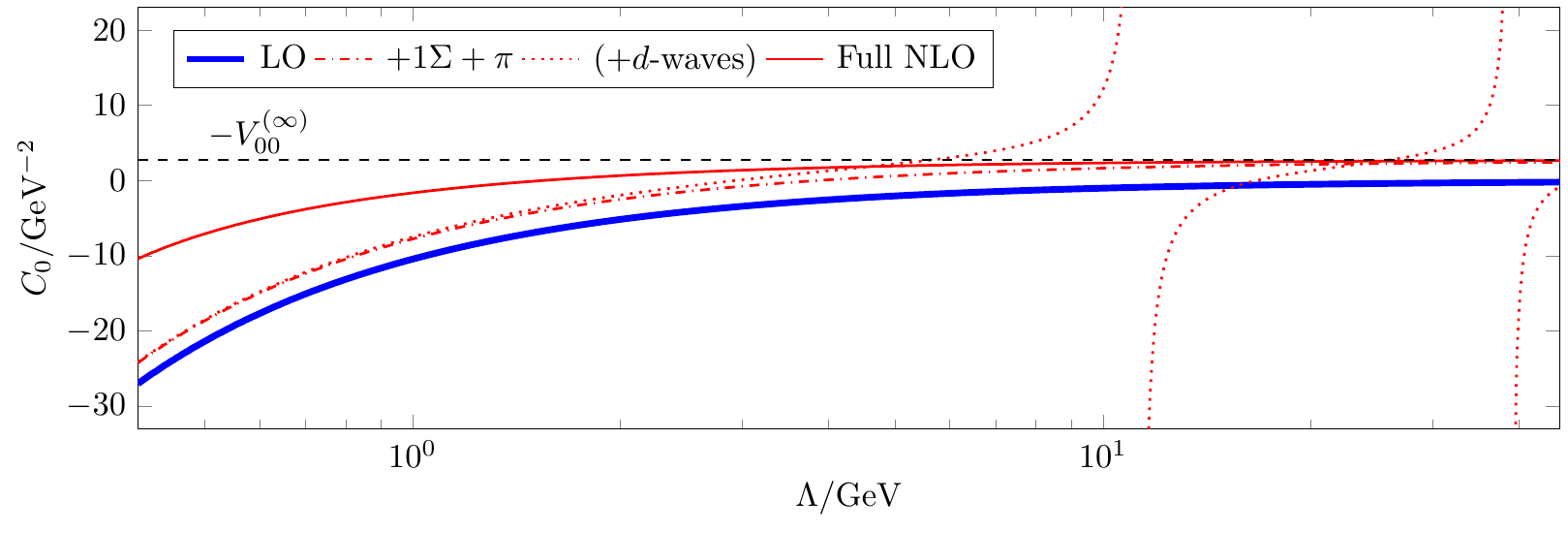}
	\end{center}
	\vspace{-1cm}
	\caption{\label{Fig:C0}Contact interaction $C_0(\Lambda)$ for $\dX=\unit[100]{keV}$.}
\end{figure}

Figure~\ref{Fig:C0} shows the curves $C_0(\Lambda)$ for $\Lambda\in\unit[{[0.4,\,50]}]{GeV}$ obtained in the different calculations. The LO result reproduces Eq.~\eqref{Eq:LOResultsGXC0}. Self-energy corrections barely influence the pole's real part and neither do they influence $C_0(\Lambda)$. In contrast, pion exchanges shift the curve by an amount $-V_{00}^{(\infty)}\approx \unit[2.72]{GeV^{-2}}$ as expected. The charged meson contribution solely suppresses parts of $C_0(\Lambda)$ that vanish as $\Lambda\rightarrow\infty$.

By including $d$-waves nonperturbatively, however, the running coupling significantly changes its signature. It exhibits consecutive singularities for fairly high cutoffs $\Lambda> \unit[11]{GeV}$, which was also observed by Baru \etal\ \cite{Baru:2011rs}. They are due to deep three-body states entering the spectrum at large cutoffs.
This kind of spectrum is a general feature of three-body systems with resonant $p$-wave interactions \cite{PhysRevLett.111.113201, Braaten:2011vf}. The deep bound states lie outside the region of validity of the EFT and do not influence the physics close to the $D\bar D^\ast$
threshold. We have explicitly checked that this is the case when we renormalize onto the shallow \X\ pole. However, in calculations resumming both $d$-wave and charged meson states at the same time, the deep bound states lead to renormalization artefacts. In particular, there are cutoffs at which no value of $C_0$ can produce the \X\ pole \cite{Baru:2011rs}. This problem is not present at NLO, where $d$-waves are negligible.

Note, that our non-perturbative $d$-wave calculation in Fig.~\ref{Fig:C0} does not correspond to a strict N${}^2$LO treatment of such contributions. Instead, one would include single $d$-wave states pertubatively, similarly to the inclusion of the charged meson loop. It remains to be seen if in such a calculation $C_0$ alone can produce the \X\ for arbitrary cutoffs.

\subsection{Line Shape of the $\bs{X(3872)}$ in $\bs{D^0\bar{D}^0\pi^0}$ Production}

\begin{figure}
	\begin{center}
		\includegraphics[scale=1]{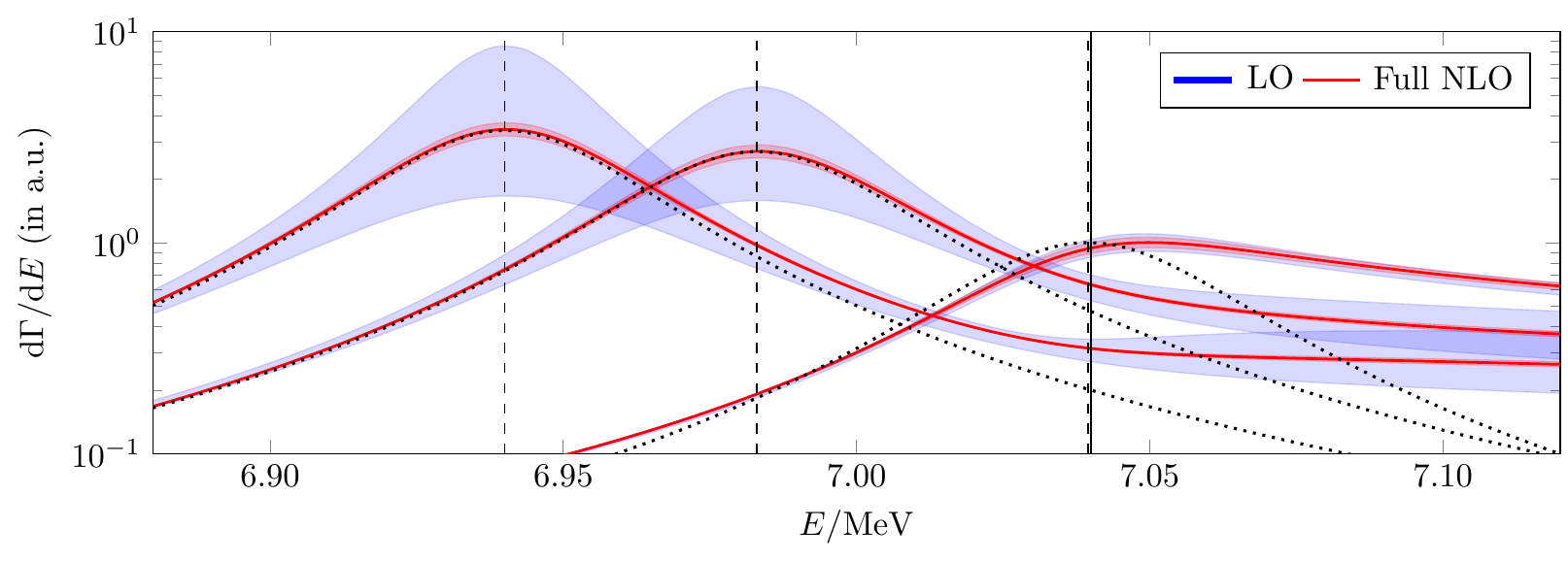}
	\end{center}
	\vspace{-1cm}
	\caption{\label{Fig:LineShape}Normalized line shapes $\text{d}\Gamma/\text{d} E$ as functions of the energy $E$ for $\dX\in\unit[\{0.5,\,57,\,100\}]{keV}$ (dashed grid lines right to left) up to NLO. The $D\bar{D}^\ast$ threshold is indicated by a solid grid line. The black dotted curves show Breit-Wigner shapes with maximum positions $\delta-\dX$ and widths $\GX^\text{(NLO)}(\dX)$. All curves are cutoff independent above the value $\Lambda=\unit[1]{GeV}$ used in the calculations.}
\end{figure}

We conclude this section by showing numerical results of the line shape of the \X\ in $D\bar{D}\pi$ production. In Fig.~\ref{Fig:LineShape}, normalized line shapes for the three values $\dX\in\unit[\{0.5,\,57,\,100\}]{keV}$ at LO and NLO are depicted. All curves are cutoff independent\footnote{Non-normalized line shapes exhibit a $\Lambda^2$-divergence, which we absorb into the short-range factor $\mathcal{F}$.} above the used value $\Lambda=\unit[1]{GeV}$. For all $\dX\geq \unit[50]{keV} \approx\, \Gast$, deviations of the peak parameters $\tilde{\delta}_X=\delta-E_\text{max}$ and $\tilde{\Gamma}_{\!X}=\text{FWHM}$ from $\dX$ and $\GX$ are negligible at NLO. Note, however, that the production rate does not possess a Breit-Wigner shape (indicated by black dotted curves). Instead, it is enhanced at the $D\bar{D}^\ast$ threshold, as observed by Braaten and Lu \cite{Braaten:2007dw}.

\begin{figure}
	\begin{center}
		\subfigure[]{
		\includegraphics[scale=1]{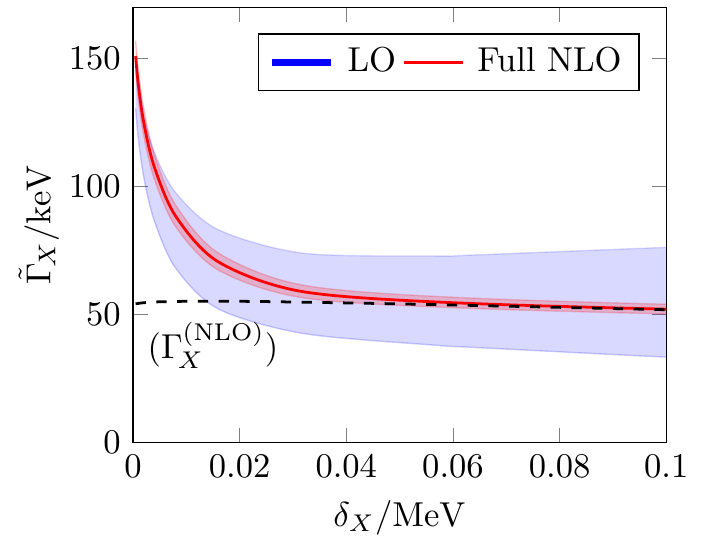}
		}
		\subfigure[]{
		\includegraphics[scale=1]{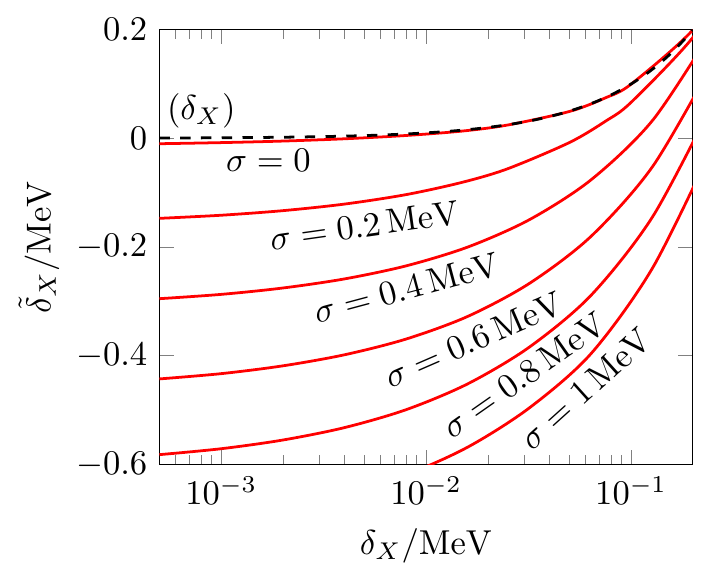}
		}
	\end{center}
	\vspace{-1cm}
	\caption{\label{Fig:PeakData}Peak parameters $\tilde{\Gamma}_{\!X}$ (a) and  $\tilde{\delta}_X$ (b) as functions of $\dX$. In (b), LO and NLO results coincide. We mimic detector resolution effects by convoluting the line shape with a Gaussian of standard deviation $\sigma$.}
\end{figure}

As $\dX$ decreases, threshold effects become more and more important. This effect can be seen in Fig.~\ref{Fig:LineShape} where the FWHM value $\tilde{\Gamma}_{\!X}$ is significantly enlarged for $\dX=\unit[0.5]{keV}$. We investigate this phenomenon in more detail in Fig.~\ref{Fig:PeakData}(a) by comparing $\tilde{\Gamma}_{\!X}$ (red solid line) to $\GX$ (black dashed line) for different $\dX$ at NLO. As soon as $\dX$ becomes comparable with $\GX\approx \unit[50]{keV}$, the line width increases up to about $\tilde{\Gamma}_{\!X}\approx \unit[150]{keV}\approx 2.8\,\GX$.
The function $\tilde{\Gamma}_{\!X}(\dX)$ turns out to be strictly monotonically decreasing. Thus, it can be inverted to determine $\dX$ from an experimentally measured line width.

\begin{figure}
	\begin{center}
		\includegraphics[scale=1]{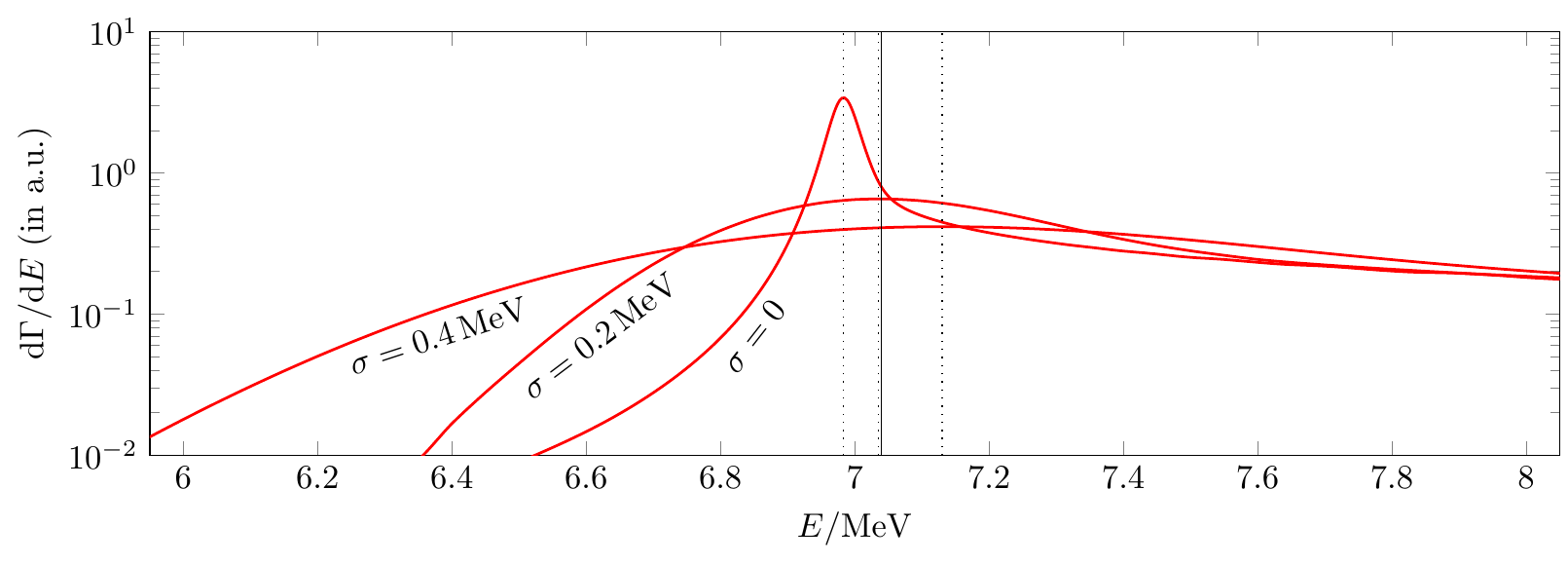}
	\end{center}
	\vspace{-1cm}
	\caption{\label{Fig:LineShapeSmeared}Smeared line shapes for $\dX=\unit[57]{keV}$ at NLO. The solid grid line represents the $D\bar{D}^\ast$ threshold, the dotted ones mark the maximum positions $E_\text{max}=\delta-\tilde{\delta}_X$.}
\end{figure}

The approximation $\tilde{\delta}_X\approx \dX$ is even valid down to $\dX\approx \unit[10]{keV}$ as shown in Fig.~\ref{Fig:PeakData}(b) (red solid lines and black dashed line, respectively). Below this value, the peak maximum crosses the $D\bar{D}^\ast$ threshold (see also Fig.~\ref{Fig:LineShape} for $\dX=\unit[0.5]{keV}$). This effect becomes even more significant if we take into account the energy resolution of the detector. We mimic its influence by convoluting the line shape with a normal distribution of standard deviation $\sigma$. Indeed, due to the threshold enhancement, the peak of the smeared line shape is shifted to higher energies, see Fig.~\ref{Fig:LineShapeSmeared}. For $\dX=\unit[57]{keV}$, a detector resolution $\sigma\geq \unit[200]{keV}$ is sufficient to shift the peak onto the threshold. Moreover, Fig.~\ref{Fig:PeakData}(b) shows that $\tilde{\delta}_X$ is almost linear in $\sigma$. This finding illustrates that in experiments the \X\ peak could occur above threshold even if the \X\ were bound.
We conclude that, in order to avoid misinterpretations of experimental findings, the detector resolution needs to be of the order of the width $\GX\approx \unit[50]{keV}$.

\subsection{Remarks on Other $\bs{X(3872)}$ Interpretations}

It should be noted, that our results only address the case of a bound \X\ state, whose pole lies below threshold ($\dX>0$) on the first Riemann sheet. Given the experimental binding energy in Eq.~\ref{Eq:dX}, the \X\ pole could in principle also lie above threshold ($\dX<0$). In this case, the molecular interpretation may not be appropriate. First of all, $s$-wave resonances cannot be produced by simple attractive potentials because they lack a centrifugal barrier \cite{Hyodo:2013iga}. In our EFT, the \X\ pole is produced by the pointlike interaction $C_0$, which does not allow for such a possibility. In Ref.~\cite{Hyodo:2013iga}, the $\Lambda_c(3595)$ was studied as a shallow $s$-wave resonance in the $\pi\,\Sigma_c$ system. It was shown that this interpretation requires an unnaturally large and negative effective range parameter, disfavoring the resonance interpretation. A similar result was obtained in Ref.~\cite{Guo:2016wpy}.

The \X\ could also be a $D\bar{D}^\ast$ virtual state on the second sheet below threshold ($\dX>0$). As shown in the zero-range approach by Braaten and Lu \cite{Braaten:2007dw}, the production rate is then given by a monotonically increasing function with maximal slope near threshold. Our theory at LO coincides with the approach by Braaten and Lu and thus it indeed allows for a virtual \X\ state as well. However, a detailed analysis of the virtual \X\ pole at NLO may require an analytic continuation of Eq.~\eqref{Eq:ScalarAmps} to the second energy sheet. Such a generalization will be part of future work.

\section{Summary and Outlook}
\label{Sec:SummaryOutlook}

In this work, we have proposed a novel EFT for the exotic \XX\ state, which can be interpreted as a loosely-bound $D^0\bar{D}^{0\ast}$ molecule in the $C=+$ channel. The EFT contains nonrelativistic $D^0$, $\bar{D}^0$ and $\pi^0$ fields and possesses exact Galilean invariance. The $D^{0\ast}$ vector meson was included as a $p$-wave resonance in the $D^0\pi^0$ sector.

Up to NLO in our power counting, we have calculated relations between the binding energy $\dX$ of the \XX, its width $\GX$, and its line shape in $D^0\bar{D}^0\pi^0$ production. For the representative value $\dX=\unit[57]{keV}$, the width is given by $\GX^\text{(NLO)}=\unit[(52.5\pm 2.2)]{keV}$. Remarkably, the corresponding uncertainty interval, stemming from experimental inputs, includes the central value of the LO result $\GX^\text{(LO)}=\unit[(53.6\pm 15.0)]{keV}$. This observation indicates a quick convergence of the theory. Moreover, the line shape exhibits a strong $D^0\bar{D}^{0\ast}$ threshold enhancement dominating the \XX\ peak for $\dX < \GX$, confirming earlier studies by Braaten and Lu \cite{Braaten:2007dw}. Our theory captures this enhancement and provides a method to systematically extract the \XX\ pole from the experimental line shape up to NLO accuracy.

Our counting is based on the characteristic momentum scales in the $D^0\pi^0$ and $D^0\bar{D}^0\pi^0$ sectors. The two-body system was analyzed in Sec.~\ref{Sec:DPi}. Exploiting Galilean invariance, we performed a comprehensive scaling analysis of $D^0\pi^0$ threshold parameters in terms of the momentum scales $\KLO= (2\mu\,\delta)^{1/2}\approx \unit[42]{MeV}$ and $\KHI\sim m_\pi\approx\unit[135]{MeV}$. As a result, the existence of the narrow $D^{0\ast}$ resonance can be explained from a single fine-tuning of QCD. It is reflected in an enhancement of the $p$-wave effective range $r_1/2\sim\KLO^{-2}\KHI^3$. Shallow $p$-wave states in other physical systems were attributed to an enhanced scattering volume $a_1$; see Refs.~\cite{Bertulani:2002sz, Bedaque:2003wa, Hammer:2011ye}. To our knowledge, the $D^{0\ast}$ is the first example of a shallow $p$-wave state, in which $a_1$ appears to be of natural size. Note, that other scaling scenarios may be possible, but they would require further fine-tunings. Radiative $D^{0\ast}$ decays were effectively included using complex interactions. At the end of the section, we derived an expansion of the full $D^{0\ast}$ propagator in the kinematic region of the \XX. The LO propagator contains the full $D^{0\ast}$ width $\Gast= \GDPi+\GDg= \unit[(53.6\pm 1.0)]{keV}$ as a constant. This ingredient is of paramount importance for the occurrence of the threshold enhancement. Propagator corrections are at least suppressed by the small ratio $\GDPi/(2\delta)\approx 0.0025$, suggesting a quick convergence of the expansion.

In Sec.~\ref{Sec:DDbarPi}, we constructed the non-perturbative $D^0\bar{D}^{0\ast}$ amplitude in the $C=+$ channel. At LO, it only contains iterations of the LO $D^{0\ast}$ propagator and the $D^0\bar{D}^{0\ast}$ contact term $C_0(\Lambda)$, which produces the \XX\ pole for arbitrary cutoffs $\Lambda$. The LO amplitude is similar to the result of Braaten and Lu \cite{Braaten:2007dw}, with the exception that the $D^{0\ast}$ width enters as a constant. The subleading nature of interactions other than $C_0$ was justified in a diagrammatic power counting. In particular, we have investigated respective loop integrals in terms of the low-momentum scales \mbox{$P_X= (2\mu_\ast |\dX|)^{1/2}\sim \unit[11]{MeV}$} and \mbox{$P_\ast= (2\mu_\ast \delta)^{1/2}\approx \unit[117]{MeV}$} of the three-body system. As a result, the $D^{0\ast}$ self-energy, $s$-wave pion exchanges and charged meson loops have to be included at NLO. This claim was verified by calculation. Higher-order $D^{0\ast}$ self-interactions, relativistic corrections, intermediate $d$-waves and charged pion exchanges can be neglected at NLO. The theory is then renormalizable for arbitrary values of the cutoff.

An important finding of this work is that for small binding energies $\dX\leq \unit[50]{keV}$, the line shape's FWHM is significantly larger than $\GX$ (up to $\approx 2.8\,\GX$). In contrast, the peak's maximum position can be described by the pole's real part even for very small binding energies, i.e., for $\dX\geq \unit[10]{keV}$. This identification, however, fails once the detector's energy resolution is taken into account. For $\dX\approx\unit[50]{keV}$, an energy resolution $\sigma>\unit[200]{keV}$ is sufficient to shift the peak maximum above the $D^0\bar{D}^{0\ast}$ threshold. This effect has to be taken into account in analyses of $D^0\bar{D}^0\pi^0$-type decays of the \XX\ \cite{Adachi:2008sua, Aubert:2007rva, Gokhroo:2006bt}. In order to not misinterpret the nature of the \XX, its peak has to be measured with a resolution of the order of the width $\GX$.

In the near future, our EFT can be used to analyze data from $D^0\bar{D}^{0\ast}$-type \XX\ decays at Belle \cite{Gokhroo:2006bt, Adachi:2008sua}. Specifically, it would be interesting to calculate the Dalitz plot for decays to $D^0\bar{D}^0\pi^0$. Moreover, we could predict the line shape of the \XX\ for production at resonance at $\bar{\text{P}}$ANDA, i.e., in processes of the type $\text{p}\bar{\text{p}}\rightarrow X(3872)\rightarrow J/\psi + X$ \cite{Prencipe:2015cgg}. Our framework could be extended in order to account for the partial widths of inelastic decay channels like $J/\psi\,\pi^+\pi^-$ by choosing a complex coupling $C_0$. However, this procedure requires a value for the branching ratio of $D^0\bar{D}^{0\ast}$-type decays of the \XX. At the moment, this quantity is only limited from below by $\unit[32]{\%}$ \cite{Gokhroo:2006bt}. Moreover, we could extend our framework to calculate line shapes for the \XX\ as a virtual $D^0\bar{D}^{0\ast}$ state.

\begin{acknowledgments}
	We thank Eric Braaten, Wael Elkamhawy, and Artem Volosniev for discussions. Moreover, we thank Eric Braaten for motivating this work by pointing out that the $D^{0\ast}$ can be introduced dynamically as a $D^0\pi^0$ resonance. This research was supported by the Deutsche Forschungsgemeinschaft through SFB 1245 ``Nuclei: From Fundamental Interactions to Structure and Stars".
\end{acknowledgments}

\appendix

\section{Calculation of the $\bs{D^{0\ast}}$ Self-Energy}
\label{Sec:Appendix2Body}

In this section, we derive the $D^\ast$ self-energy function $\Sigma$ as depicted in Fig.~\ref{Fig:FeynDyson}. Moreover, we show that in both the MS and PDS schemes, the sign $\Delta_1=\pm 1$ of Eq.~\eqref{Eq:LDPi} is positive.

Let $p^\mu=(p^0,\bs{p})$ be the total $D^\ast$ four-momentum. Due to Galilean symmetry, the bare self-energy can only depend on the center-of-mass energy $\Ecm=p^0-\bs{p}^2/(2M)$. For incoming and outgoing polarizations $i,j\in\{1,\,2,\,3\}$, it reads
\begin{align}
	\label{Eq:SelfEnergyIntegral1}
	-i\,\Sigma_{ij}^\text{(b)}(\Ecm)
	=\ &\int\!\!\frac{\text{d}^4 l}{(2\pi)^4}
	\frac{i\,(g\,l_i)}{(\alpha p^0-l^0) - \frac{\left(\alpha\bs{p}-\bs{l}\right)^2}{2m_D}+i\epsilon}\,
	\frac{i\,(-g\,l_j)}{((1-\alpha)p^0+l^0) - \frac{\left((1-\alpha)\bs{p}+\bs{l}\right)^2}{2m_\pi}+i\epsilon}
	\\
	\label{Eq:SelfEnergyIntegral2}
	=\ &i\,g^2
	\int\!\frac{\text{d}^3 l}{(2\pi)^3}
	\frac{l_i\,l_j}{\frac{\bs{l}^2}{2\mu}+\frac{\bs{p}^2}{2M}-p^0-i\epsilon}
	\\
	=\ &i\,g^2\,2\mu
	\int\!\frac{\text{d}^3 l}{(2\pi)^3}
	\frac{\bs{l}^2\,\delta_{ij}/3}{\bs{l}^2-2\mu(\Ecm+i\epsilon)}
	\\
	\equiv\ &-i\,\Sigma^\text{(b)}(\Ecm)\,\delta_{ij}\,.
\end{align}
In Eq.~\eqref{Eq:SelfEnergyIntegral1}, we have made explicit use of Galilean symmetry in taking the relative $D\pi$ four-momentum $l^\mu\equiv (l^0,\,\bs{l})\equiv \alpha l_\pi^\mu-(1-\alpha)l_D^\mu$ with $\alpha= m_D/(m_\pi+m_D)$ as a loop integration variable. The $l^0$ integral has been performed using the residue theorem. Moreover, the integral in Eq.~\eqref{Eq:SelfEnergyIntegral2} vanishes for $i\ne j$ (asymmetric under $l_i\rightarrow -l_i$) and is otherwise independent of $i$. Therefore, we may replace $l_i\,l_j\rightarrow \bs{l}^2\,\delta_{ij}/3$.

In order to calculate the right-hand integral in Eq.~\eqref{Eq:SelfEnergyIntegral2}, we turn to $d$ spatial dimensions and introduce a subtraction scale $\Lambda_\text{PDS}$. We find
\begin{align}
	-i\,\Sigma^\text{(b)}(\Ecm)
	=\ &i\,g^2\,\frac{2\mu}{3}\left(\frac{\Lambda_\text{PDS}}{2}\right)^{3-d}
	\int\!\frac{\text{d}^d l}{(2\pi)^d}
	\frac{\bs{l}^2}{\bs{l}^2-2\mu(\Ecm+i\epsilon)}
	\\
	\label{Eq:SelfEnergyIntegral3}
	=\ &i\,g^2\,\frac{2\mu}{3}\left(\frac{\Lambda_\text{PDS}}{2}\right)^{3-d}
	\frac{d}{2}\frac{\Gamma(-d/2)}{(4\pi)^{d/2}}\left[-2\mu(\Ecm+i\epsilon)\right]^{d/2},
\end{align}
which has a pole in $d=2$ but not in $d=3$. In the MS scheme, we evaluate Eq.~\eqref{Eq:SelfEnergyIntegral3} for $d=3$ yielding the expression given in Eq.~\eqref{Eq:Sigma}.

However, it is enlightening to take a look at the result in the PDS scheme in which poles in $d=2$ are subtracted as well \cite{Kaplan:1998tg}. For this purpose, we introduce the counterterm
\begin{align}
	\label{Eq:SelfEnergyIntegral4}
	-\Delta\Sigma_{ij}^\text{(PDS)}(\Ecm)
	\equiv\ &-i\,g^2 \delta_{ij} \frac{\mu}{6\pi}\,\frac{\Lambda_\text{PDS}}
	{d-2}\left[-2\mu(\Ecm+i\epsilon)\right]
	\\
	\equiv\ &-i\,\Delta\Sigma^\text{(PDS)}(\Ecm)\,\delta_{ij}\,,
\end{align}
which vanishes for $\Lambda_\text{PDS}=0$. In this limit, we can easily recover the MS result. The full PDS result for $\Sigma$ is then given by
\begin{align}
	\Sigma^\text{(PDS)}(\Ecm)\equiv\ 
	&\Sigma^\text{(b)}(\Ecm)+\Delta\Sigma^\text{(PDS)}(\Ecm)
	\Big|_{d=3}
	\\
	=\ &-g^2\,\frac{\mu}{6\pi}\left(
		\Lambda_\text{PDS}\,2\mu(\Ecm+i\epsilon)+\left[-2\mu(\Ecm+i\epsilon)\right]^{3/2}
	\right).
\end{align}
For a general $\Delta_1=\pm 1$, the $p$-wave effective range now reads
\begin{equation}	\label{Eq:IdentitiesGenMu}
	\frac{r_1}{2}=-\left(\frac{6\pi}{\mu}\,\frac{\Delta_1}{2\mu g^2} + \Lambda_\text{PDS}\right),
\end{equation}
while the scattering volume $a_1$ and all higher-order parameters $\mathcal{P}_{2n}$ are independent of $\Lambda_\text{PDS}$.
From Eq.~\eqref{Eq:IdentitiesGenMu}, it is obvious that two threshold parameters, i.e., $a_1^{-1}$ and $r_1/2$, are needed in $D\pi$ scattering. Using a finite momentum cutoff $\lambda$, the term $\Lambda_\text{PDS}$ would correspond to a linear divergence in $\lambda$. To take care of this linear divergence, the (bare) parameter $r_1$ can not be chosen as zero. Moreover, a cubic divergence
in $\lambda$ would enter in $a_1^{-1}$.

If we neglect radiative decays of the $D^\ast$, the EFT Lagrangian must be Hermitian, implying $g^2>0$. Furthermore, we know that $r_1<0$. In the MS scheme, Eq.~\eqref{Eq:IdentitiesGenMu} tells us that $\Delta_1=+1$. The same is true in the PDS scheme for a subtraction point $0\leq \Lambda_\text{PDS}<-r_1/2$. This choice is reasonable since $|r_1/2|\approx \unit[17.1]{GeV}$ is much larger than the expected breakdown scale $\KHI\sim m_\pi$. Thus, the $D^\ast$ is a physical particle in our theory.

\section{Determination of the $\bs{D^{0\ast}\leftrightarrow D^0\pi^0}$ Coupling}
\label{Sec:AppendixCoupling}

We infer a value for $g^2$ from the well-known decay widths of the charged $D^{+\,\ast}$ meson using isospin symmetry. The $D^{+\,\ast}$, similar to the $D^{0\,\ast}$, represents a $p$-wave resonance of constituents $D^+\pi^0$ or $D^0\pi^+$. Its total width for pionic decays is given by $\Gc=\unit[82(2)]{keV}$.
The experimental masses of the charged scalar mesons read \mbox{$m_{D^+}= \unit[1869.58(9)]{MeV}$} and
\mbox{$m_{\pi^+}= \unit[139.57018(35)]{MeV}$} \cite{PDG2017}. Moreover, the mass differences in the charged channels, \mbox{$\delta_{+0}\equiv m_{D^{+\ast}}-m_{D^0}-m_{\pi^+}
=\unit[5.855(2)]{MeV}
$} and
\mbox{$\delta_{++}\equiv m_{D^{+\ast}}-m_{D^+}-m_{\pi^0}
=\unit[5.69(8)]{MeV}
$}, are again much smaller than the particle masses but much larger then $\Gc/2$. We see that the charged channels exhibit scale separations comparable to the neutral case. The couplings of the transitions \mbox{$D^{+\ast}\rightarrow D^0\pi^+$} and \mbox{$D^{+\ast}\rightarrow D^+\pi^0$} are given by $2g$ and $g$, respectively. This is a consequence of isospin symmetry \cite{Braaten:2015tga}.

We assume higher-order parameters in the $D^{+\ast}$ to scale naturally. Therefore, Eq.~\eqref{Eq:GammaGCorrelation} can be modified for the charged channels by writing
\begin{equation}	\label{Eq:ChargedDecays}
	\Gc/2= 
	\left(
		|\Sigma(\delta_{+0})|\Big|_{\substack{
			\mu\rightarrow\mu_{+0}
			\\
			g^2\rightarrow 2g^2
		}}
		+|\Sigma(\delta_{++})|\Big|_{\substack{
			\mu\rightarrow\mu_{++}
			\\
			g^2\rightarrow g^2
		}}
	\right)\Big(1+\mathcal{O}\left(\chi\,\kappa\right)\Big)
\end{equation}
with $\mu_{+0}\equiv\left(m_{D^0}^{-1}+m_{\pi^+}^{-1}\right)^{-1}
$ and $\mu_{++}\equiv\left(m_{D^+}^{-1}+m_{\pi^0}^{-1}\right)^{-1}
$. This yields
\begin{equation}
	\label{Eq:gSquared}
	g^2=\frac{3\pi}{\sqrt{2}}\ \frac{
		\Gc/2
	}{
		2\mu_{+0}^{5/2}\delta_{+0}^{3/2}
		+\mu_{++}^{5/2}\delta_{++}^{3/2}
	}
	\Big(1+\mathcal{O}\left(\chi\,\kappa\right)\Big)
	= \unit[3.48(8)\cdot 10^{-8}]{MeV^{-3}}\,.
\end{equation}

\section{Relativistic Corrections}
\label{Sec:AppendixRelativistics}

In order to estimate the influence of relativistic corrections, we equip $D$ and $\pi$ with exact Klein-Gordon propagators. Let $\varrho^\mu$ be the relativistic four-momentum and $p^0=\varrho^0-m_a$ with $a\in\{D,\,\pi\}$ the kinetic energy of the respective meson. We can then write the propagators in the form
\begin{equation}
	i\,G_a(\varrho^\mu) = i\left[\varrho^\mu \varrho_\mu-m_a^2\right]^{-1}
	=\frac{1}{2m_a}\,i\,\left[
		p^0-\frac{\bs{p}^2}{2m_a}+\frac{(p^0)^2}{2m_a}
	\right]^{-1},
	\quad a\in\{D,\,\pi\}\,.
\end{equation}
For the pion case, this propagator can be described by the kinetic Lagrangian term
\begin{equation}
	\mathcal{L}_{\text{kin},\,\pi} = 2m_\pi\,\pi^\dagger\left[
		i\,\partial_0+\frac{\nabla^2}{2m_\pi}
		-\frac{\partial_0^2}{2m_\pi}
	\right] \pi\,.
\end{equation}
After field redefinitions $\pi^{(\dagger)}\rightarrow \pi^{(\dagger)}(2m_\pi)^{1/2}$, we recover the nonrelativistic Lagrangian of Eq.~\eqref{Eq:LKin} if the term quadratic in $\partial_0$ is neglected. Thus, this term represents the relativistic correction to the respective one-body propagator.

This finding allows us to estimate corrections from relativistic pion exchanges. Let $\bs{p}_\text{in/out}$ be the incoming/outgoing relative $D\bar{D}^\ast$ momentum. Then the kinetic energy of the exchanged pion is given by $p^0=E-(p_\text{in}^2+p_\text{out}^2)/(2m_D)$. For both low-momentum scales $p_\text{in/out}\sim P_X\sim\unit[11]{MeV}$ and $p_\text{in/out}\sim P_\ast=\unit[117]{MeV}$ in the three-body sector, this energy lies in the range $[0,E]$ and thus $p^0\leq E\sim \delta$. We see that relativistic corrections in exchanged pions are suppressed by a factor $p^0/(2m_\pi)\leq \delta/(2m_\pi)\sim 0.5\,\kappa^2$. Thus, they do not enter before N${}^2$LO.

For the estimation of relativistic corrections in the $D^\ast$ propagator, we investigate a $D\pi$ pair moving at a total kinetic energy energy $p^0$ and a total momentum $\bs{p}$. As in the two-nucleon case~\cite{Chen:1999tn}, Lorentz invariance ensures that $p^0$ and $\bs{p}$ are related to the center-of-mass kinetic energy $p^0_\text{cm}$ via
\begin{equation}
	\label{Eq:LorentzInvariance}
	p^0-\frac{\bs{p}^2}{2M}+\frac{(p^0)^2}{2M}
	=p^0_\text{cm}+\frac{(p^0_\text{cm})^2}{2M}
\end{equation}
with $M=m_D+m_\pi$.
The $D^\ast$ pole position appears at $(p^0_\text{cm})^\text{(pole)}=E_\ast$. By plugging this condition into Eq.~\eqref{Eq:LorentzInvariance} and using $\bs{p}^2\ll |M+E_\ast|^2$, we determine the pole position in the general frame to be
\begin{equation}
	(p^0)^\text{(pole)}=E_\ast+\frac{\bs{p}^2}{2(M+E_\ast)}
	-\frac{\bs{p}^4}{8(M+E_\ast)^3}+\cdots\,.
\end{equation}
The full $D^\ast$ propagator can then be written like
\begin{equation}
	i\,G_\ast(p^\mu)=\frac{i\,Z(p^0)}{p^0-(p^0)^\text{(pole)}}+\text{reg}\,,
\end{equation}
with $p^0=E-\bs{p}^2/(2m_D)+\bs{p}^4/(8m_D^3)-\cdots$ in the $D\bar{D}^\ast$ system. In the nonrelativistic limit, the difference $p^0-(p^0)^\text{(pole)}$ has to recover the Galilean-invariant expression $\Ecm-E_\ast=E-\bs{p}^2/(2\mu_\ast)-E_\ast$ frequently used in this paper. Indeed, we obtain this expression by further expanding at $E_\ast/M\approx 0$, yielding
\begin{equation}
	p^0-(p^0)^\text{(pole)}
	=E-\frac{\bs{p}^2}{2\mu_\ast}\Bigg(
		1
		-\underbrace{
			\frac{2\mu_\ast E_\ast}{M^2}
		}_{\sim E_\ast/M}
		-\underbrace{
			\frac{2\mu_\ast\bs{p}^2}{8M^3}
			+\frac{2\mu_\ast\bs{p}^2}{8m_D^3}
		}_{\sim \bs{p}^2/M^2}
		+\dots
	\Bigg)-E_\ast\,,
\end{equation}
where we used $2\mu_\ast\sim M$. All the corrections in the parentheses are suppressed by the total $D\pi$ mass and thus extremely small. The first one is comparable to $\kappa^5+i\,\kappa^{10}$ while the second and third one are of order $\kappa^6$. Since only imaginary corrections contribute to the width, relativistic corrections in the $D^\ast$ propagator only enter at N${}^5$LO.

\section{Partial Wave Projection}
\label{Sec:AppendixPartWaveComps}

We absorb angular dependences of the amplitude into vector spherical harmonics\footnote{Note that our definition differs by a factor $\sqrt{4\pi}$ from the one used in Ref.~\cite{Baru:2011rs}.}
\begin{equation}
	\boldsymbol{Y}_{[L,\,1]Jm_J}(\boldsymbol{n})
	\equiv
	\sqrt{4\pi}\sum_{m_L,\,m_S}
	\Big\langle
		L\ m_L;\ 1\ m_S\ 
		\Big|\ 
		[L,\,1]\ J\ m_J
	\Big\rangle\ 
	Y_{L}^{m_L}(\boldsymbol{n})\,\bs{\chi}_{m_S}\,.
\end{equation}
The function $Y_{L}^{m_L}(\bs{n})$ denotes a spherical harmonic evaluated at a unity vector $\bs{n}$, while $\bs{\chi}_{m_S}$ is a spherical basis vector in $\mathbb{C}^3$. With $\bs{e}_p\equiv \bs{p}/p$ and $\bs{e}_{p'}\equiv \bs{p}'/{p'}$, the expansion for the amplitude reads
\begin{equation}
	\label{Eq:PartWaveExp}
	T^{\,ij}\left(\bs{p},\,\bs{p}';\,E\right)
	\equiv \sum_J \sum_{L,L'}\, T_{L L';\, J}\left(p,\,p';\,E\right)
	\sum_{m_J}
	\left(
		\bs{Y}_{[L,\,1]Jm_J}\left(\bs{e}_p\right)
	\right)^{i}
	\,
	\left(\bs{Y}_{[L',\,1]Jm_J}^\ast
	\left(\bs{e}_{p'}\right)\right)^{j}\,.
\end{equation}
The $m_J$-sum over the two vector spherical harmonics in Eq.~\eqref{Eq:PartWaveExp} yields projection operators of the form $P_{LL';\,J}^{ij}$.
In the same fashion, we expand the pion exchange potential $V^{ij}$.

The \X\ appears in the $J=1$ channel with $L,L'\in\{0,\,2\}$. The relevant components of the pion exchange potential read
\begin{align}	\label{Eq:PWPot00}
	V_{00;\,1}\left(p,\,q;\,E\right)
	=\ &-\frac{1}{6}\,g^2 m_\pi 
	\left[
		\alpha\left(p^2+q^2\right)I_0
		+\left(\alpha^2+1\right)pq\, I_1
	\right]\left(p,\,q;\,E\right),
	\\		\label{Eq:PWPot02}
	V_{02;\,1}\left(p,\,q;\,E\right)
	=\ &\frac{\sqrt{2}}{6}\,g^2 m_\pi 
	\left[
		\alpha\,q^2\,I_0
		+ \left(\alpha^2+1\right)pq\,I_1
		+ \alpha\,p^2\,I_2
	\right]\left(p,\,q;\,E\right),
	\\	\label{Eq:PWPot20}
	V_{20;\,1}\left(p,\,q;\,E\right)
	=\ &V_{02}\left(q,\,p;\,E\right),
	\\	\label{Eq:PWPot22}
	V_{22;\,1}\left(p,\,q;\,E\right)
	=\ &-\frac{1}{3}\,g^2 m_\pi 
	\left[
		\left(\alpha^2+\frac{1}{10}\right)pq\,I_1
		+\alpha\left(p^2+q^2\right)I_2
		+\frac{9}{10}\,pq\,I_3
	\right]\left(p,\,q;\,E\right).
\end{align}
They involve integrals
\begin{equation}
	\label{Eq:In}
	I_l\left(p,\,q;\,E\right) \equiv
	\int_{-1}^1\,\text{d}x\,\frac{P_l(x)}{
		\frac{1}{2\alpha}\left(p^2+q^2\right)
	-m_\pi \left( E + i\epsilon \right)+pq\,x
	}\quad (l\geq 0)
\end{equation}
over Legendre polynomials $P_l$.

\section{Charged Meson Propagators in the Vicinity of the $\bs{X(3872)}$}
\label{Sec:AppendixCharged}

Like the neutral mesons, their charged partners $D^\pm$ and $\pi^\pm$ can be treated nonrelativistically in the energy region of the \X. This can be seen from the fact that all charged three-body thresholds, i.e., $D^\pm D^\mp \pi^0$, $D^0 D^- \pi^+$ and $\bar{D}^0 D^+ \pi^-$, lie closer to the \X\ than the neutral one; see Fig.~\ref{Fig:MassSketch}. The respective propagators,
\begin{align}
	i\,G_{D^\pm}(p^\mu)=&\ 
	i\left[p^0-\frac{\bs{p}^2}{2m_{D^+}}-(m_{D^+}-m_{D^0})\right]^{-1},
	\\
	i\,G_{\pi^\pm}(p^\mu)=&\ 
	i\left[p^0-\frac{\bs{p}^2}{2m_{\pi^+}}-(m_{\pi^+}-m_{\pi^0})\right]^{-1},
\end{align}
take care of the mass differences between charged and neutral partners. Note that all charged three-body thresholds lie above the \X. Therefore, they are completely off shell for energies close to the \X\ pole.

Charged vector mesons $D^{\pm\ast}$ can be constructed as $p$-wave resonances of their constituents with resonance energies of the order $\delta_{+0},\,\delta_{++}\lesssim \delta$ (see Appendix~\ref{Sec:AppendixCoupling} for numerical values). Moreover, their self-energies are also suppressed by a factor of order $\chi$ compared to the resonance energies. For this reason, the $D\pi$ threshold power counting developed in Sec.~\ref{Sec:DPi} can be applied to the charged resonances. This means that self-energies and higher-order corrections are sub-leading for small energies. In the region of the \X\ , we may take the $D^{\pm\ast}$ propagators to be
\begin{equation}
	i\,G_{D^{\pm\ast}}(p^\mu)
	=i\left[
		p^0-\frac{\bs{p}^2}{2m_{D^{+\ast}}}-({m_{D^{+\ast}}-m_{D^{0\ast}}})
	\right]^{-1}.
\end{equation}

Similar to the neutral $D^{0\ast}$, we could in principle introduce a constant decay width $\Gamma\left[D^{+\ast}\rightarrow D^+\gamma\right]=\unit[1.3(4)]{keV}$ in the $D^{+\ast}$ propagator \cite{PDG2017}. As a result, we would have to replace $\nu\rightarrow \nu-i\,\Gamma\left[D^{+\ast}\rightarrow D^+\gamma\right]/2$ in Eq.~\eqref{Eq:ChargedInteraction} and $\Gast\rightarrow \Gast-\Gamma\left[D^{+\ast}\rightarrow D^+\gamma\right]$ in Eq.~\eqref{Eq:ChargedCorrection}. This tiny modification is negligible at NLO.

\bibliography{thresholdX_v3.bbl}

\end{document}